\pdfoutput=1
\documentclass[11pt]{article}

\usepackage{graphicx}
\usepackage{amsmath}
\usepackage{amssymb}
\usepackage{dcolumn}
\usepackage{bm}
\usepackage{slashed}
\usepackage{color}
\usepackage{authblk}
\usepackage{ulem}
\usepackage{CJK}
\usepackage{hyperref}
\usepackage[numbers,sort&compress]{natbib}

\def\l{{\lambda}}
\def\L{{\Lambda}}
\def\d{{\delta}}
\def\D{{\Delta}}
\def\o{{\omega}}
\def\O{{\Omega}}
\def\e{{\epsilon}}

\def\a{{\alpha}}
\def\b{{\beta}}

\def\f{{\phi}}
\def\g{{\gamma}}
\def\G{{\Gamma}}

\def\h{\eta}
\def\p{{\pi}}
\def\P{{\Pi}}
\def\m{{\mu}}
\def\n{{\nu}}
\def\r{{\rho}}
\def\s{{\sigma}}
\def\S{{\Sigma}}
\def\t{{\tau}}
\def\th{{\theta}}

\def\ps{{\psi}}

\def\x{{\xi}}
\def\P{{\Pi}}

\def\({\left(}
\def\){\right)}
\def\[{\left[}
\def\]{\right]}
\def\jp{{J/\ps}}
\def\pmh{\tilde{p}}

\newcommand{\pd}{{\partial}}
\newcommand{\dg}{\dagger}
\newcommand{\pr}{\parallel}

\newcommand{\tr}{\text{tr}}
\newcommand{\pp}{\perp}

\newcommand{\fctrq}{g^2N_fT_F}
\newcommand{\fctrg}{g^2C_A}
\newcommand{\rts}{\sqrt{s_{NN}}}

\date{\today}

\begin{document}

\begin{CJK}{UTF8}{gbsn}

\title{\bf Spin alignment of quarkonia in vortical quark-gluon plasma}
%{\cfundlink}

\author[1]{{Yuhao Liang(梁宇浩)}
\thanks{liangyh83@mail2.sysu.edu.cn}}
%\affil[1]{School of Physics and Astronomy, Sun Yat-sen University, Zhuhai 519082, China}
\author[1]{{Shu Lin(林树)}
\thanks{linshu8@mail.sysu.edu.cn}}
\affil[1]{School of Physics and Astronomy, Sun Yat-sen University, Zhuhai 519082, China}

\maketitle

%\address{3)}{}

\begin{abstract}
The spin alignment of $J/\psi$ with respect to event plane in relativistic heavy ion collisions exhibits a significant signal. We propose a possible mechanism for spin alignment through spin dependent dissociation of quarkonia in a vortical quark-gluon plasma. The spin dependent dissociation is realized through inelastic scattering between constituents of quarkonium and those of quark-gluon plasma polarized by the vorticity. The spin dependent dissociation rate is found to depend on the directions of vorticity, quantization axis, and quark momentum. We implement our results in a dissociation dominated evolution model for quarkonia in the Bjorken flow, finding the spin $0$ state is slightly suppressed compared to the average of the other two, which is consistent with the sign found in experiments. We also find absence of logarithmic enhancement in binding energy in the vortical correction to dissociation rate, which is understood from the requirement that a spin dependent dissociation can only come from quark coupling to a pair of chromomagnetic and chromoelectric field.
\end{abstract}
\hspace{2em}Keywords: spin alignment,quarkonium,quark-gluon plasma

\newpage

\section{Introduction}

The spin physics in relativistic heavy ion collisions (HIC) has become an emergent field under intensive studies recently. It was first realized that the final particles in off-central heavy ion collisions should be spin polarized \cite{Liang:2004ph}. The first measurement was made in global polarizaiton of $\L$ hyperons \cite{STAR:2017ckg}, which has been nicely understood from the coupling of $\L$ hyeprons spin with vorticity of quark-gluon plasma (QGP) \cite{Becattini:2013fla,Fang:2016vpj,Li:2017slc}. Further measurement of local polarization has further revealed more refined structure of spin interaction \cite{STAR:2019erd}, allowing us to characterize the spin response to more general hydrodynamic gradients \cite{Fu:2021pok,Becattini:2021iol,Yi:2021ryh,Wu:2022mkr,Lin:2022tma,Lin:2024zik,Wang:2024lis,Fang:2024vds}.

A closely related phenomenon is the spin alignment of vector meson, originally proposed from polarization of the two constituent quarks based on quark model \cite{Liang:2004xn,Yang:2017sdk}. In HIC, the spin alignment is measured with a quantization axis chosen to be perpendicular to the event plane. The spin component along the axis defines $1,\,-1,\,0$ states. With the polarization of quarks from spin-vorticity coupling, we would have a spin alignment quadratic in vorticity. However, the measurement of spin alignment of $\f$ meson has revealed significantly large signal, with the probability of finding $0$ state significantly larger than the other two state. A variety of mechanisms have been explored to understand the large signal \cite{Sheng:2022wsy,Sheng:2023urn,Kumar:2023ghs,Yang:2024qpy,Wagner:2022gza,DeMoura:2023jzz,Xu:2024kdh,Sheng:2024kgg,Fu:2023qht,Zhao:2024ipr}\footnote{Besides that, we can also refer to "F. Li , S.Y.F. Liu, Tensor Polarization and Spectral Properties of Vector Meson in QCD Medium. (2022)"}, see \cite{Chen:2024afy} for a recent review. More recently there has been a proposal on constraining the mechanisms experimentally \cite{Lv:2024uev}.

A second vector meson observed with spin alignment is the $J/\ps$. The measurement has found a spin alignment different from the counterpart of $\f$ meson in sign \cite{STAR:2022fan}. This might not be a surprise as the production mechanism of $J/\ps$ is completely different from that of $\f$, which comes mainly from thermal production. Since the charm quark is much larger than the temperature of QGP, thermal production is negligible. $J/\ps$ can be produced either directly in initial hard scatterings or from recombination of charm and anti-charm quarks, which are also produced by initial hard scatterings. Both sources are subject to significant medium modification. The early produced $J/\ps$ partly dissociate due to interaction with the medium and the charm quarks evolving with the medium also carries information about the medium before they recombine to form $J/\ps$. We shall refer to the two cases loosely as dissociation and recombination production respectively. The dissociation and recombination production are known to dominate in low and high energy collisions respectively \cite{Braun-Munzinger:2000csl,Zhao:2007hh,Zhou:2014kka}.

While the vorticity coupling to spin of constituent quarks does give the correct sign for spin alignment of $\jp$, the overall magnitude is expected to be much smaller than what has been observed in experiments \cite{Liang:2004xn,Yang:2017sdk}. This is because the spin alignment is quadratic in the phenomenologically small vorticity. In fact, this is not the only effect of vorticity on spin alignment. In this paper, we will propose another possible mechanism through spin dependent dissociation in a QGP with vorticity. Since dissociation occurs throughout evolution of the QGP, the small vorticity can be compensated by long evolution time. For simplicity, we focus on dissociation production in medium-high energy collisions. We consider dissociation rate of quarkonia in a spinning QGP characterized by vorticity $\o$. For quarkonia in the vortical QGP, we expect by rotational symmetry that the vortical contribution to dissociation rate for quarkonia in spin $s$ state can be parameterized by $\G_s=\G_{1}s\widehat{\boldsymbol{p}}\cdot\boldsymbol{\omega}(\widehat{\boldsymbol{n}}\cdot\widehat{\boldsymbol{p}})+\G_{2}s\widehat{\boldsymbol{n}}\cdot\boldsymbol{\omega}+O(\o^2)$, with $\hat{n}$ and $\hat{p}$ being the direction of quantization axis and quarkonium momentum. $s=1,-1,0$ labels the spin states of the quarkonia. The survival probability of the initially produced quarkonia is given by exponential decay as $\exp(-\int\G_s dt)$. Since exponential decay is a concave function, the splitting in the dissociation rate above leads to a suppressed rate of $0$ state compared to the average of the other two. This is independent on the sign of $\G_1$ and $\G_2$ in the parameterization above. The goal of this paper is to work out the spin dependent part of the dissociation rate and study its impact on spin alignment of $\jp$.

The paper is structured as follows: in Section~\ref{sec_diss} we first review the spin independent dissociation of quarkonia in QGP in the quasi-free picture, and then calculate spin dependent vortical correction to dissociation rate of quarkonia. Apart from dependence on spin component $s$, the correction is also found to depend on quantization axis and momentum of quarkonia; in Section~\ref{sec_pheno}, we apply the results to calculate the spin alignment of $\jp$ in medium-high energy collisions and discuss the phenomenological implications; Section~\ref{sec_concl} is devoted to conclusion and outlook. Details of calculations are reserved to three appendices. 

\section{Quarkonia dissociation in vortical QGP}\label{sec_diss}

With application to $\jp$ in mind, we consider dissociation of spin-triplet S-wave quarkonium states (to be referred to as quarkonium state in short) in vortical QGP. Since magnitude of vorticity produced in HIC is $\sim10\,\text{MeV}$ still much smaller than typical temperature of QGP $\sim200\,\text{MeV}$, we may treat vorticity dependent part of dissociation rate as a perturbation. The dissociation rate of quarkonium state in the absence of vorticity has been well understood, which we briefly review below.

\subsection{Quarkonia dissociation rate in equilibrium QGP}

The dissociation of an unthermalized quarkonia arise from two types of processes: gluo-dissociation and inelastic scattering. In the former, the color singlet quarkonium dissolves into a color octet by absorbing a gluon from QGP \cite{Bhanot:1979vb,Peskin:1979va,Song:2005yd,Chen:2017jje}. This process is leading order in coupling constant. In the latter, one of the heavy (anti-)quark constituents scatters with light quark or gluon in the QGP, converting the color singlet quarkonium into color octet. It is next to leading order in coupling constant and formally suppressed. However, the naive power counting ignores the bound state nature of quarkonium. In fact the gluo-dissociation process is only possible for bound state: in the limit of vanishing binding energy, the gluo-dissociation cross section simply vanishes by vanishing phase space. Therefore it is crucial to take into account the effect of finite binding energy. Indeed it has been realized that the cross section for gluo-dissociation is rather small as quarkonium is loosely bounded at high temperature \cite{Karsch:1987pv}. It follows that the dominant process for dissociation is the inelastic scattering. A quasi-free picture has been proposed for quarkonium, in which the two constituents of quarkonium scatter with light quarks and gluons in QGP independently \cite{Grandchamp:2001pf,Grandchamp:2002wp}, see also \cite{Chen:2018dqg,Zhao:2024gxt}. The validity of the scenario is studied in framework of potential non-relativistic QCD \cite{Brambilla:2011sg,Brambilla:2013dpa,Brambilla:2010vq,Yao:2018sgn,Chen:2025mrf}, which treats bound state effect more systematically.

In fact, binding energy also play a crucial role in the inelastic scattering process, which includes Coulomb scattering and Compton scattering. It is well known that the perturbative damping rate for heavy quark suffers from infrared (IR) divergence when the exchanged gluon carries very soft momenta \cite{Peigne:2008nd,Braaten:1991we}. Fortunately the finite binding energy to be denoted as $\e_b$ sets a threshold for the energy of exchanged gluon to dissolve the quarkonium, thus the binding energy effectively cuts off the IR divergence, leading to a logarithmic enhancement in the binding energy. Similar IR divergence does not occur in Compton scattering as the exchanged particle is the heavy constituent \cite{Peigne:2008nd}. At sufficient small binding energy, the Coulomb scattering process is logarithmically enhanced giving the dominant contribution to the cross section. Such an enhancement of the form $\ln\frac{T}{\e_b}$ has been found in \cite{Brambilla:2010vq}, with $T$ being temperature of the QGP $T$. Focusing on the logarithmically enhanced contribution allows one to consider Coulomb scattering only.

With Coulomb scattering, the evaluation of dissociation rate can be reduced to the effective one-loop self-energy diagram for heavy quark shown in Fig.~\ref{fig1}. The gluon propagator running in the loop is the hard thermal loop (HTL) resummed one, which screens the IR divergence of Coulomb scattering. The remaining logarithmic divergence is cut off by $\e_b$ as we already discussed.

\subsection{Vorticity correction to dissociation rate}

The question we now ask is how the vorticity enters the diagram above. We still assume the quarkonium to be unthermalized so that the vorticity only affects the light quarks and gluons. The modifications are known from quantum kinetic theories, see \cite{Hidaka:2022dmn} for a recent review. Crucially the vorticity modifies only the on-shell light quarks and gluons but not their off-shell counterparts. Aiming at vortical correction to dissociation rate to leading logarithmic order $\ln\frac{T}{\e_b}$, we will also restrict ourselves to Coulomb scattering process. The role played by the vorticity is to affect the distribution and polarization of light degrees of freedom in the initial and final states. 

With the light degrees of freedom polarized by vorticity, we expect a spin-dependent damping rate of heavy quarks inside the quarkonium though collisions with polarized light degrees of freedom. The spin-dependent damping rate for the constituent can further lead to spin-dependent dissociation rate for the bound state. The spin averaged damping rate can be related to self-energy as \cite{Bellac:2011kqa}
\begin{align}\label{G_rep}
	\G=\frac{1}{4E_p}\tr[({\slashed P}+m)\S^>(P)].
\end{align}
%In fact $\S^>$ should be understood as imaginary part of retarded self-energy $\S_R$ as 
%\begin{align}
%\S^>(P)=2\text{Im}[\S_R(P)](1+f(P\cdot u))=2\text{Im}[\S_R(P)],
%\end{align}
%where the first equality follows from Kubo-Martin-Schwinger relation and the second equality holds for a probe (unthermalized) heavy quark. 
To derive damping rate for a particular spin state, we need to project the self-energy into a spin state. To find the appropriate spin projection operator, we first recall the following representation of free retarded propagator for particle state
\begin{align}
{S_R(P)=\frac{1}{2E_p}\sum_s\frac{u_s(P)\bar{u}_s(P)}{p_0-E_p+i\e}=\frac{1}{2E_p}\sum_s\frac{u_s(P){u}^\dg_s(P)\g^0}{p_0-E_p+i\e}},
\end{align}
where $u_s(P)$ is the eigenspinor solution to the free Dirac equation normalized as $u^\dg_r u_s=2E_p\d_{rs}$. In fact, this is general eigenstate representation of retarded function in quantum mechanics applied to Dirac theory with labels of eigenstates being spin. We readily identify  {$\frac{u_s(P){u}^\dg_s(P)}{2E_p}$} as the projection operator onto a specific spin state and {$\frac{\sum_s u_s(P){u}^\dg_s(P)}{2E_p}$}is the identity operator in Dirac space. With the same logic, we can rewrite the spin average in \eqref{G_rep} more explicitly as
\begin{align}\label{G_rep2}
{\G=\frac{1}{2}\tr[\frac{\sum_s u_s(P){u}^\dg_s(P)\g^0}{2E_p}\S^>(P)]}.
\end{align}
This suggests the following damping rate for specific spin state
\begin{align}\label{Gamma_s}
{\G_s=\frac{1}{2E_p}\tr[u_s(P){u}^\dg_s(P)\g^0\S^>(P)]=\frac{1}{2E_p}\tr[u_s(P)\bar{u}_s(P)\S^>(P)]}.
\end{align}
Let us comment on the validity of this representation: \eqref{Gamma_s} implicitly assumes $E_p$ to be degenerate and free eigenspinors to be used. In principle, both come from solution to the following Dirac equation in the presence of self-energy
\begin{align}
\({\slashed P}-m-\S_R\)u_s(P)=0,
\end{align}
with $\S_R$ being retarded self-energy. However in perturbation theory the $\S_R\sim \S^>$ is small and consists leading order contribution to $\G_s$, so that we may ignore modifications to energy and eigenspinor in \eqref{Gamma_s}, which are higher order effect.

The case of anti-particle can be worked out similarly. Since the vorticity effect in a charge neutral QGP is expected to be identical for particle and anti-particle, we will focus on damping rate for particle below. 

%By rotational symmetry, we may parameterize the vortical correction to damping rate as
%\begin{align}
%\begin{equation}\label{Gamma_para}
%\Gamma_{s}=\G_{1}(p,p^{0})s\widehat{\boldsymbol{p}}\cdot\boldsymbol{\omega}(\widehat{\boldsymbol{n}}\cdot\widehat{\boldsymbol{p}})+\G_{2}(p,p^{0})s\widehat{\boldsymbol{n}}\cdot\boldsymbol{\omega},
%\end{equation}
%\end{align}
%where $\hat{\bm{n}}$ and $\hat{\bm{p}}$ are directional vector for quantization axis and heavy quark momentum. $s=\pm1/2$ is the projection of spin along the quantization axis.

\begin{figure}[h]
	\centering
	\includegraphics[width=0.45\linewidth]{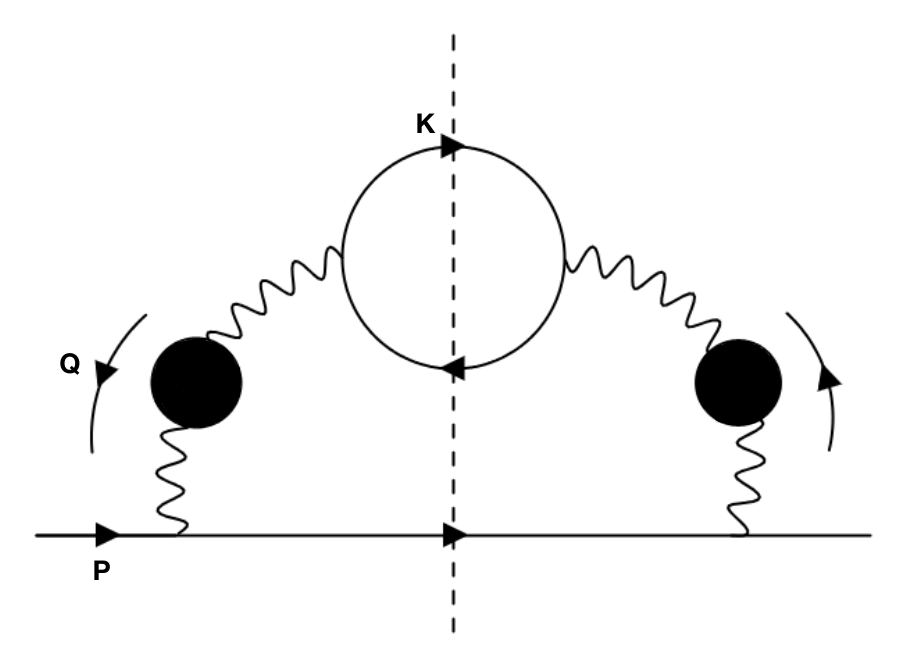}
	\includegraphics[width=0.45\linewidth]{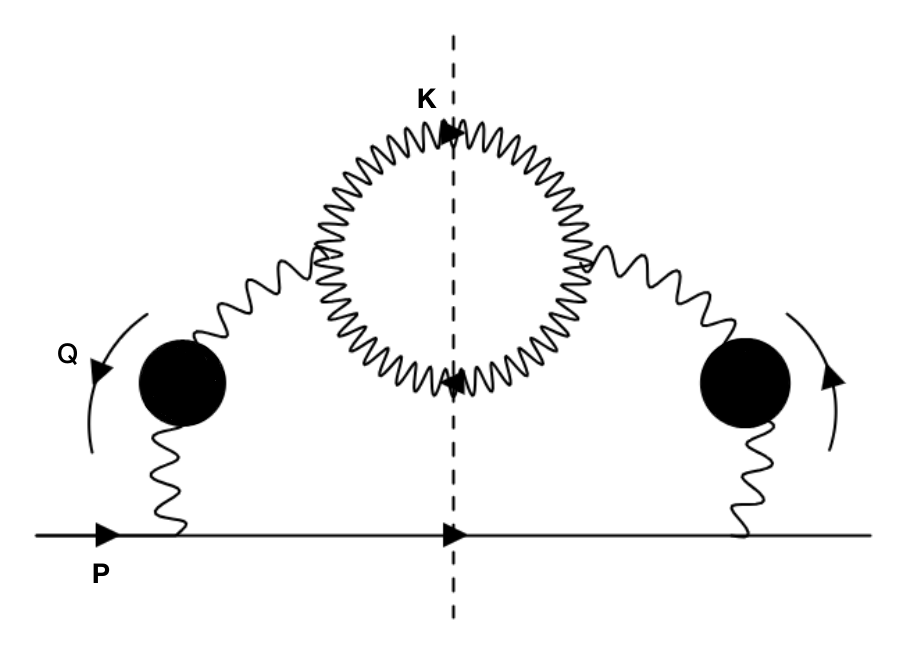}
	\caption{The self-energy diagrams for heavy quark. The cutted propagators can be put on-shell. The left/right diagram corresponds to Coulomb scattering off light quarks/gluons. The logarithmic enhancement comes from exchange of gluons with soft momenta, so that we need to use HTL resummed propagators for them, indicated by blobs.}\label{fig1}
\end{figure}

Now we are ready to calculate self-energy of heavy quark in a vortical QGP. As argued in the previous subsection, to the leading logarithmic order in $\e_b$, only Coulomb scattering contributes, corresponding to the self-energy diagrams in Fig.~\ref{fig1}. 

%\begin{figure}[h]
%	\centering
%	\includegraphics[width=0.4\linewidth]{image1.png}
%	\includegraphics[width=0.4\linewidth]{image2.png}
%	\includegraphics[width=0.4\linewidth]{image3.png}
%	\caption{In the above figure,gamma=$g^4f_1$, and the figure shows the dependence of gamma on the parameters $\e$, $m$, and $T$.}
%	\label{fig:enter-label}
%\end{figure}
The vorticity corrections to the diagrams enter through modifications of propagators for light quarks and gluons. The explicit forms are known from quantum kinetic theories. The gluon lesser propagator in the Coulomb gauge is given by \cite{Huang:2020kik,Hattori:2020gqh,Lin:2021mvw}
%
%To obtain the first-order correction to the dissociation  rate with respect to vorticity, we need the off-equilibrium propagator  in vortical plasma,we can find this  in the reference\cite{}. \\
%For gluon correction propagator:
\begin{align}\label{gluon_prop}
&D_{\m\r}^<(X,Q)=D_{\mu\rho}^{<(0)}(X,Q)+D_{\mu\rho}^{<(1)}(X,Q),\\
&D_{\m\r}^{<(0)}(X,Q)=2\p P_{\m\r}^T\d(Q^2)\e(Q\cdot u)f(Q\cdot u),\nonumber\\
&D_{\mu\rho}^{<(1)}(X,Q)=2\pi\epsilon(Q\cdot u)\delta(Q^2)\left[\frac{iP_{\mu\lambda}Q^\lambda P_{\rho\sigma}P^{\sigma\beta}}{2{(Q\cdot u)}^2}\partial_\beta f(Q\cdot u)-(\mu\leftrightarrow\rho)-i\frac{\epsilon_{\mu\rho\alpha\beta}Q^\alpha u^\beta}{{(Q\cdot u)}^2}Q^\nu\omega_\nu f'(Q\cdot u)\right],\nonumber
\end{align}
where $X$ is a coarse-grained coordinate labeling fluid elements. We assume uniform temperature and slow-varying fluid velocity $u_\m(X)$. The vorticity is defined as $\o^\m=\frac{1}{2}\e^{\m\n\r\s}u_\n\pd_\r u_\s$. $P^{\m\r}_T=-\h^{\m\r}+\frac{Q^\m u^\r+Q^\r u^\m}{Q\cdot u}-\frac{Q^\m Q^\r}{(Q\cdot u)^2}$ and $P_{\m\n}=u_\m u_\n-\h_{\m\n}$ are transverse and spatial projectors respectively.$\e$ is the sign function,$f$ is the Bose-Einstein distribution function. The superscripts $(0)$ and $(1)$ indicate the order in vorticity. $D_{\m\r}^{<(1)}$ can be further simplified. In the fluid rest frame with $u^\m=(1,0,0,0)$, only spatial components of $D_{\m\r}^{<(1)}$ are nonvanishing and read
\begin{align}
D_{ij}^{<(1)}&=2\p\d(Q^2)\e(q_0)\big[-\frac{i q_i q_k\pd_j u_k}{2q_0^2}f'-(i\leftrightarrow j)-\frac{i\e^{ijk}q_k}{q_0^2}(-q_l\o_l)f'\big]\nonumber\\
&=2\p\d(Q^2)\e(q_0)\big[\frac{i\e^{ijk}q_kq_l\o_l}{2q_0^2}+\frac{i\e^{ijk}\o_k}{2}\big]f',
\end{align}
where $f'\equiv\frac{\pd}{\pd q_0}f(q_0)$. We have used $\pd_ju_k=\e^{jkl}\o_l$ in a vortical fluid and the Schouten identity $-q_i\e^{jkl}+q_j\e^{kli}-q_k\e^{lij}+q_l\e^{ijk}=0$ in the second equality.
The greater propagator $D_{\m\r}^>$ can be obtained by the replacement $f\to 1+f$ in \eqref{gluon_prop}. Note that $D_{\m\r}^{<(1)}=D_{\m\r}^{>(1)}$, indicating the spectral function is not modified at this order. The color structure is the trivial one inherited from the free gluon propagator, which we have suppressed for notational simplicity.
%for gluon correction propagator and
%For fermion correction propagator:

The quark lesser propagator is given by \cite{Hidaka:2017auj,Gao:2018jsi}
\begin{align}\label{quark_prop}
&S^<(X,K)=S^{<(0)}(X,K)+S^{<(1)}(X,K),\\
&S^{<(0)}(X,K)=-2\p\e(K\cdot u)\d(K^2){\slashed K}\widetilde{f}(K\cdot u),\nonumber\\
    &\textit{S}^{<(1)}(X,K)=-\pi K^{\mu}\widetilde{\Omega}_{\mu\nu}\gamma^{\nu}\gamma^{5}\e(K\cdot u)\delta(K^{2})\widetilde{f}^{\prime}(K\cdot u),
\end{align}
with $\tilde{\Omega}^{\mu\nu}=\omega^\mu u^\nu-\omega^\nu u^\mu$. $\widetilde{f}$ is the Fermi-Dirac distribution function. The greater propagator $S^>$ can be obtained from \eqref{quark_prop} by the replacement $-\widetilde{f}\to1-\widetilde{f}$. Again we have $S^{<(1)}=S^{>(1)}$, indicating no modification of spectral function at this order. Below we will work in rest frame of fluid with $u^\m=(1,0,0,0)$ and suppress the explicit dependence on $X$.
%for fermion correction propagator.
%Here $P^{\mu\nu}=-g^{\mu\nu}+u^{\mu}u^{\nu}$,$\tilde{\Omega}^{\mu\nu}=\omega^\mu u^\nu-\omega^\nu u^\mu$,$f^{\prime}(k^{0})=\frac{\partial f}{\partial (\beta k^{0})}$，$\widetilde{f}'(k_{0})=\frac{\partial \widetilde{f}}{\partial (\beta k^{0})}$.\\
%For the quark dissociation rate, we can use the following expression \cite{}: 

%We focus on vorticity correction to the damping rate
%\begin{align}
%    \Gamma_s^{(1)}=\frac{1}{E}tr[u_s(P)\bar{u}_s(P)\Sigma^{>(1)}(P)]
%\end{align}
%When considering the case of Coulomb scattering, $\Sigma^>(p)$ in the above equation can be represented by Figure 1\cite{}:
The self-diagrams adopt the same representation below
\begin{align}\label{Sigma}
    \Sigma^{>(1)}(P)&=g^{2}C_F\int_{Q}\gamma^{\mu}S_H^{>}(P+Q)\gamma^{\nu}D_{\nu\mu}^{<(1)}(Q)\notag\\&=-g^{2}C_F\int_{Q}\gamma^{\mu}S_H^{>}(P+Q)\gamma^{\nu}D_{\nu\alpha}^{R}(Q)\Pi^{\alpha\beta<(1)}(Q)D_{\beta\mu}^{A}(Q),
\end{align}
with $C_F=\frac{N_c^2-1}{2N_c}$ coming from color sum $t_{ij}^at^a_{jk}=C_F\d_{ik}$.
Here $S_H^>(P+Q)=2\p({\slashed P}+{\slashed Q}+m)\d((P+Q)^2-m^2)$ is the greater propagator for probe heavy quark. $D_{\nu\mu}^{<(1)}(Q)=-D_{\nu\alpha}^{R}(Q)\Pi^{\alpha\beta<(1)}(Q)D_{\beta\mu}^{A}(Q)$ is the vortical correction to gluon propagator, with the correction sits entirely in the gluon self-energy $\P_{\a\b}^{<(1)}$ \cite{Hou:2020mqp}. The gluon self-energy arise from either quark loop or gluon loop captured by the two diagrams in Fig.~\ref{fig1}respectively. $D_{\m\n}^{R/A}$ are retarded/advanced gluon propagators in the absence of vorticity, given by
\begin{align}\label{DRA}
D_{\mu\nu}^{R}=u_\mu u_\nu\Delta_L+P_{\mu\nu}^T\Delta_T,\quad D_{\mu\nu}^A=D_{\mu\nu}^R{}^*.
\end{align}
$P^T_{\m\n}$ is the transverse projector defined before and $u^\m u^\n$ is the longitudinal projector in Coulomb gauge. We use HTL resummed propagators relevant for exchanged gluon with soft momenta \cite{Bellac:2011kqa}, for which
\begin{align}
&\D_T=\frac{-1}{Q^2-m_g^2\(x^2+\frac{x(1-x^2)}{2}\ln\frac{x+1}{x-1}\)},\nonumber\\
&\D_L=\frac{-1}{q^2+2m_g^2\(1-\frac{x}{2}\ln\frac{x+1}{x-1}\)}.
\end{align}
$m_g$ is the gluon thermal mass defined by $m_g^2=\frac{1}{6}g^2T^2\(C_A+\frac{1}{2}N_f\)$. $x=\frac{q_0}{q}$ with $q_0$ and $q$ being temporal and spatial components of momentum in QGP frame.
Using \eqref{quark_prop}, we obtain $\P^{\a\b<(1)}$ from quark loop is given by
\begin{align}\label{Pi_quark}
&\Pi^{\alpha\beta<(1)}_q(Q)=-2g^2N_{f}T_F\int_Ktr\bigg[\gamma^\alpha S^{<(1)}(K+Q)\gamma^\beta S^{>(0)}(K)\bigg]\\&=-4ig^2N_fT_F (2\pi)^{2}\int_{K}\epsilon^{\alpha\nu\beta\lambda}(K+Q)^{\mu}\widetilde{\Omega}_{\mu\nu}K_{\lambda}\epsilon(k^{0})\epsilon(k^{0}+q^{0})\left(1-\widetilde{f}\left (k_{0}\right)\right)\widetilde{f}^{\prime}(k_{0}+q_{0})\delta(K^{2})\delta\left(\left(K+Q\right)^{2}\right)\nonumber,
\end{align}
with $T_F=\frac{1}{2}$ coming from the color sum $\tr[t^at^b]=T_F\d^{ab}$ and $N_f$ being number of light quark flavors. The prefactor $2$ arises because vortical correction can enter either quark propagator in the loop. Clearly, $\P^{\a\b<(1)}$ is anti-symmetric indices. Using \eqref{gluon_prop}, we obtain $\P^{\a\b<(1)}$ from gluon loop as
\begin{align}\label{Pi_gluon}
&\Pi^{\alpha\beta<(1)}_g(Q)=-g^2C_A\frac{1}{2}\int_K\[D_{\mu\rho}^{<(0)}(K+Q)D_{\sigma\nu}^{>(1)}(K)+ D_{\mu\rho}^{<(1)}(K+Q)D_{\sigma\nu}^{>(0)}(K)\]\\
&\Big[g^{\mu\alpha}(-K-2Q)^\nu+g^{\alpha\nu}(Q-K)^\mu+g^{\nu\mu}(2K+Q)^\alpha\Big] \Big[g^{\rho\beta}(K+2Q)^\sigma+g^{\beta\sigma}(K-Q)^\rho+g^{\sigma\rho}(-2K-Q)^\beta\Big]\nonumber,
\end{align}
where $C_A=N_c$ comes from color sum $f^{acd}f^{bcd}=C_A\d^{ab}$. The two terms in the square bracket in the first line correspond to vortical corrections to two gluon propagators. A symmetry factor $1/2$ is included. We now show it is also anti-symmetric in indices. By relabeling of the indices $\m\leftrightarrow\s$, $\n\leftrightarrow\r$ and redefinition of momentum $K\to -K-Q$, which amounts to interchange of two vertices, we find
\begin{align}\label{Pi_gluon2}
&\Pi^{\alpha\beta<(1)}_g(Q)=-g^2C_A\frac{1}{2}\int_K\[D_{\s\n}^{<(0)}(-K)D_{\m\r}^{>(1)}(-K-Q)+ D_{\s\n}^{<(1)}(-K)D_{\m\r}^{>(0)}(-K-Q)\]\\
&\Big[g^{\s\alpha}(K-Q)^\r+g^{\alpha\r}(K+2Q)^\s+g^{\r\s}(-2K-Q)^\alpha\Big] \Big[g^{\n\beta}(Q-K)^\m+g^{\beta\m}(-K-2Q)^\n+g^{\m\n}(2K+Q)^\beta\Big].
\end{align}
Using the properties $D_{\mu\rho}^{<(0)}(P)=D_{\mu\rho}^{>(0)}(-P)$, and $D_{\s\n}^{>(1)}(P)=-D_{\s\n}^{<(1)}(-P)$, we immediately find %the term $D_{\m\r}^{<(0)}(K+Q)D_{\s\n}^{>(1)}(K)$ in \eqref{Pi_gluon} is opposite to the term $D_{\s\n}^{<(1)}(-K)D_{\m\r}^{>(0)}(-K-Q)$ in \eqref{Pi_gluon2} and same conclusion holds for the other term in the square brackets, so that we have
 $\P^{\a\b<(1)}_g=-\P^{\b\a<(1)}_g$. Moreover, $\P^{\a\b<(1)}$ is purely imaginary from explicit representations \eqref{Pi_quark}, \eqref{Pi_gluon} and \eqref{gluon_prop}. It follows that $D_{\n\m}^{<(1)}$ is also anti-symmetric indices and purely imaginary. The anti-symmetric property of $D_{\n\m}^{<(1)}$ allows for the following replacement in the product of gamma matrices in \eqref{Sigma}
\begin{align}\label{gamma_product}
&\gamma ^{\mu } \gamma ^{\l } \gamma ^{\nu }=\gamma ^{\nu } \eta ^{\mu \lambda }-\gamma ^{\lambda }\eta ^{\mu \nu }+\gamma ^{\mu } \eta ^{\lambda \nu } -i  \epsilon ^{\mu \lambda \nu \rho }\gamma ^5\gamma _{\rho }\to  -i  \epsilon ^{\mu \lambda \nu \rho }\gamma ^5\gamma _{\rho },\nonumber\\
&\gamma ^{\mu } \gamma ^{\nu }=\eta ^{\mu \nu }-i \Sigma ^{\mu \nu }\to -i \Sigma ^{\mu \nu },
\end{align}
with $\S^{\m\n}=\frac{i}{2}[\g^\m,\g^\n]$.

%Plugging \eqref{DRA} into \eqref{Sigma}, and using anti-symmetric property of $\P^{\a\b<(1)}$, we have 
%\begin{align*}
%& \Sigma^{>}(X,P)=g^{2}C_F2\pi\int_{Q}\delta\left(\left(P+Q\right)^{2}-m^{2}\right)\Bigg[\textcolor{red}{\gamma^{i}({\slashed P}+{\slashed Q}+m)\gamma^{j}P_{i\overline{\alpha}}^{T,Q}P_{\overline{\beta}j}^{T,Q}\Pi^{\overline{\alpha}\overline{\beta}<(1)}(Q)|\Delta_T|^2}\\&+\textcolor{blue}{\bigg(\gamma^{0}\left({\slashed P}+{\slashed Q}+m\right)\gamma^{i}P_{i\overline{\beta}}^{T,Q}\Pi^{0\overline{\beta}<(1)}+\gamma^{i}({\slashed P}+{\slashed Q}+m)\gamma^{0}P_{i\overline{\alpha}}^{T,Q}\Pi^{\overline{\alpha}0<(1)}\bigg)\frac{1}{2}(\Delta_L\Delta_T^*+\Delta_T\Delta_L^*)}\Bigg].
%\end{align*}
%The contractions of gluon self-energy are evaluated in appendix~\ref{} with the following results
%\begin{align}\label{contractions}
%ww
%\end{align}

To evaluate the trace in \eqref{Gamma_s}, we need the following representation of eigenspinors
\begin{align}
\mathrm{u_s(P)} = \sqrt{\frac{p_0+m}{2}}
\begin{pmatrix}
\left(1-\frac{\vec{p}\cdot\vec{\sigma}}{p_0+m}\right)\xi_s \\
\left(1+\frac{\vec{p}\cdot\vec{\sigma}}{p_0+m}\right)\xi_s
\end{pmatrix}.
\end{align}
$\x_s(s=+/-)$ are spin up/down spinor along a given quantization axis $\widehat{\boldsymbol{n}}$, with the following explicit expressions
\begin{align}
\xi_{+}=\frac{1}{\sqrt{2(1-\widehat{n}_{z})}}\binom{\widehat{n}_{x}-i\widehat{n}_{y}}{1-\widehat{n}_{z}},\quad\xi_-=\xi_{+}(\bm{n}\to-\bm{n}).%\xi_{-}=\frac{1}{\sqrt{2(1+\widehat{n}_{z})}}\binom{\widehat{n}_{x}-i\widehat{n}_{y}}{-1-\widehat{n}_{z}}
\end{align}
We shall work out the case $s=+$ below. The other case can be simply obtained by flipping the direction of $\rm{n}$. The product $u_s(P)\bar{u}_s(P)$ is worked out as
\begin{align}
u_+(P)\bar{u}_+(P)&=\frac{p_0+m}{4}\Bigg[(1-\pmh^2)I+\frac{1}{2}\S^{ij}\e^{ijk}\(n_k(1+\pmh^2)-2\pmh\cdot\hat{n}\pmh_k\)+2\S^{0i}\e^{ijk}\pmh_j\hat{n_k}\nonumber\\
&+(1+\pmh^2)\g^0+2\pmh\cdot\hat{n}\g^5\g^0-2\pmh_i\g^i-\(\hat{n}_i(1-\pmh^2)+2\pmh\cdot\hat{n}\pmh_i\)\g^5\g^i\Bigg],
\end{align}
where we have defined $\tilde{p}=\frac{\vec{p}}{p_0+m}$. Using the replacement \eqref{gamma_product}, we evaluate the trace as
\begin{align}\label{trace}
&\tr[u_+(P)\bar{u}_+(P)\g^\m({\slashed P}+{\slashed Q})\g^\n D_{\n\m}^{<(1)}(Q)]\nonumber\\
&=-(p_0+m)\big[-i\e^{\m\l\n0}(P+Q)_\l D_{\n\m}^{<(1)}(Q)2\pmh\cdot\hat{n}-i\e^{\m\l\n i}(P+Q)_\l D_{\n\m}^{<(1)}(Q)\(n_i(1-\pmh^2)+2\pmh\cdot\hat{n}\pmh_i\)\big],\nonumber\\
&\tr[u_+(P)\bar{u}_+(P)(-i\S^{\m\n})D_{\n\m}^{<(1)}(Q)m]\nonumber\\
&=-(p_0+m)\big[-2iD_{i0}^{<(1)}(Q)2m\e^{ijk}\pmh_j n_k-iD_{ji}^{<(1)}(Q)m\e^{ijk}\(n_k(1+\pmh^2)-2\pmh\cdot\hat{n}\pmh_k\)\big].
\end{align}
%
%\begin{align*}
%\mathrm{u_{s}~(P)~\overline{u}_{s}~(P)~\simeq m~
%	\begin{pmatrix}
%	{\xi_{s}~\xi_{s}}^{\dagger} & {\xi_{s}~\xi_{s}}^{\dagger} \\
%	{\xi_{s}~\xi_{s}}^{\dagger} & {\xi_{s}~\xi_{s}}^{\dagger}
%	\end{pmatrix}~\equiv m~
%	\begin{pmatrix}
%	{\sigma_{s}} & {\sigma_{s}} \\
%	{\sigma_{s}} & {\sigma_{s}}
%	\end{pmatrix}}
%\end{align*}
%$\xi_s$ determines the spin state,we choose $\widehat{\boldsymbol{n}}$ as quantization axis.We are interested in the splitting between spin up/down (s=+/-), which correspond to:
%\begin{align*}
%\xi_{+}=\frac{1}{\sqrt{2(1-\widehat{n}_{z})}}\binom{\widehat{n}_{x}-i\widehat{n}_{y}}{1-\widehat{n}_{z}},\xi_{-}=\frac{1}{\sqrt{2(1+\widehat{n}_{z})}}\binom{\widehat{n}_{x}-i\widehat{n}_{y}}{-1-\widehat{n}_{z}}
%\end{align*}
Using $D_{\nu\mu}^{<(1)}(Q)=-D_{\nu\alpha}^{R}(Q)\Pi^{\alpha\beta<(1)}(Q)D_{\beta\mu}^{A}(Q)$, anti-symmetric property of $\P^{\a\b<(1)}$ and explicit representation of $D^{R/A}$ \eqref{DRA}, we can see only the following contractions of gluon self-energy are needed: $P_T^{m'm}\P^{0m<(1)}$ and $P^{m'm}_TP^{n'n}_T\P^{mn<(1)}$. These have been worked out in appendix~\ref{sec_app_A}, with the following results
\begin{align}\label{contractions_Pi}
&P_T^{m'm}\P^{0m<(1)}_q=-2\fctrq\int_K\big[4i\e^{m'jk}(k+q)_0\o_j\hat{q}_k k_\pr\big](2\p)^2\d((K+Q)^2)\d(K^2)\widetilde{f}'(k_0+q_0)\(1-\widetilde{f}(k_0)\),\nonumber\\
&P^{m'm}_TP^{n'n}_T\P^{mn<(1)}_q=-2\fctrq\int_K\big[-4i\e^{m'n'k}\hat{q}_k\hat{q}_l\o_l\((k^\pr+q) k^\pr-(k+q)_0k_0\)\big]\notag\\&\times(2\p)^2\d((K+Q)^2)\d(K^2)\widetilde{f}'(k_0+q_0)\(1-\widetilde{f}(k_0)\)\nonumber\\
&P_T^{m'm}\P^{0m<(1)}_g=-\fctrg\int_K(2k+q)_02q_l\Bigg[\(2-\frac{(k_\pr+q)q}{(k_0+q_0)^2}\)\frac{i\e^{lm'a}k_\pp^2\o_a^\pp}{4k_0^2}+\(2+\frac{k_\pr}{2q}-\frac{k_\pp^2+2(k_\pr+q)^2}{4(k_0+q_0)^2}\)\nonumber\\
&\times\frac{i\e^{lm'a}\o_a^\pp}{2}\Bigg](2\p)^2\d((K+Q)^2)\d(K^2)f(k_0+q_0)f'(k_0),\nonumber\\
&P_T^{m'm}P_T^{n'n}\P^{mn<(1)}_g=-\fctrg\int_K\Bigg[i\e^{n'lm'}\o_l^\pr\(\frac{k_\pp^2k_\pr q}{2k_0^2}+\frac{k_\pp^2}{4}\)\(-4\frac{(k_\pr+q)q}{(k_0+q_0)^2}-4\)+4q\frac{k_\pp^2k_\pr}{(k_0+q_0)^2}(-k_\pp^2)\nonumber\\
&\times\frac{i\e^{km'n'}\o_k^\pr}{2k_0^2}+\frac{4q k_\pr k_\pp^2}{(k_0+q_0)^2}\(\frac{i\e^{n'm'a}\o_a^\pr k_\pr^2}{2k_0^2}+\frac{i\e^{n'm'a}\o_a^\pr}{2}\)\Bigg](2\p)^2\d((K+Q)^2)\d(K^2)f(k_0+q_0)f'(k_0).
\end{align}
with the subscripts ``q'' and ``g'' denoting contributions from quark and gluon loops respectively. The common part of the phase space integration is calculated as
\begin{align}\label{phase_space}
&\int_K\d(K^2)\d((K+Q)^2)=\int k^2dkd\cos\th d\f\frac{1}{2k}|_{k_0=k}+\int k^2dkd\cos\th d\f\frac{1}{2k}|_{k_0=-k}\nonumber\\
&=\int_{\frac{q-q_0}{2}} k^2dk 2\p\frac{1}{2k}\frac{1}{2kq}|_{k_0=k}+\int_{\frac{q+q_0}{2}} k^2dkd\cos\th d\f\frac{1}{2k}|_{k_0=-k},
\end{align}
where $\int d\cos\th\d((K+Q)^2)=\frac{1}{2kq}$ fixing $\cos\th=\frac{q_0^2-q^2+2k_0q_0}{2kq}$.

Using \eqref{Gamma_s}, \eqref{Sigma} and \eqref{trace}, and performing angular average of $q$ (details reserved for appendix~\ref{sec_app_B}, we obtain the following representation for $\G_s^{(1)}$
\begin{equation}\label{Gamma12}
   \Gamma_{s}^{(1)}=\G_{1}s\widehat{\boldsymbol{p}}\cdot\boldsymbol{\omega}(\widehat{\boldsymbol{n}}\cdot\widehat{\boldsymbol{p}})+\G_{2}s\widehat{\boldsymbol{n}}\cdot\boldsymbol{\omega}.
\end{equation}
In the above, $s=\pm\frac{1}{2}$ has been inserted for the cases of both spin states. The functions $\G_1$ and $\G_2$ are scalar functions that depend on the quark momentum $p$, energy $p_0$, binding energy $\e_b$ and temperature $T$, with explicit forms given by:
\begin{align}\label{Gamma_1}
 \G_1=&g^2\frac{N_c^2-1}{2N_c}\frac{p_0+m}{4p_0}2\pi i\int\frac{dq_0}{(2\pi)^3}dq\frac{q^2}{2pq}\Bigg[4|\Delta_T|^2\(A_1^{q}+A_1^{g}\)\frac{1}{q^2}\bigg[(1-\tilde{p}^2)q_0(3q_\parallel^2-q^2)\notag\\&+4\tilde{p}^2q_\parallel^2(p_0+q_0+m)-\notag4\tilde{p}(pq_\parallel^2+q_\parallel q^2)\bigg]+4\frac{\Delta_L\Delta_T^*+\Delta_T\Delta_L^*}{2}\(A_2^{q}+A_2^{g}\)\frac{1}{q}\notag\\&\times\bigg[(\tilde{p}^2-1)(2pq_\parallel+3q_\parallel^2-q^2)+4\tilde{p}^2(q^2-q_\parallel^2)+4m\tilde{p}q_\parallel\bigg]\Bigg]
 \end{align}
 \begin{align}\label{Gamma_2}
\G_2&=g^2\frac{N_c^2-1}{2N_c}\frac{p_0+m}{4p_0}2\pi i\int\frac{dq_0}{(2\pi)^3}dq\frac{q^2}{2pq}\Bigg[4|\Delta_T|^2\(A_1^{q}+A_1^{g}\)\frac{1}{q^2}(1-\tilde{p}^2)q_0(q^2-q_\parallel^2)\notag\\&+4\frac{\Delta_L\Delta_T^*+\Delta_T\Delta_L^*}{2}\(A_2^{q}+A_2^{g}\)\frac{1}{q}\bigg[\left(1-\tilde{p}^2\right)\left(2pq_\parallel+q^2+q_\parallel^2\right)-4m\tilde{p}q_\parallel\bigg]\Bigg]
\end{align}
Here %$\tilde{p}=\frac{p}{p_0+m}$,
$q_\parallel\equiv q\cdot\hat{p}=\frac{q_0^2-q^2+2p_0q_0}{2p}$ following from $\d((P+Q)^2-m^2)=\d(2P\cdot Q+Q^2)$ for on-shell $P$. The integration bounds of $q_0$ and $q$ are given respectively by $q_0\in[\e_b,+\infty)$, $q\in[\sqrt{p^2+2p_{0}q_{0}+q_{0}^{2}}-p  , \sqrt{p^2+2p_{0}q_{0}+q_{0}^{2}}+p]$. The functions $A_1^{q/g}$ and $A^{q/g}_2(q^0,q)$ are related to the projected gluon self-energies in \eqref{contractions_Pi} as
\begin{align}\label{A12_def}
    &P^{m'm}_TP^{n'n}_T\P^{mn<(1)}_{q/g}=\e^{m'n'l}\o_\pr^l A_1^{q/g},\nonumber\\
   &P_T^{m'm}\P^{0m<(1)}_{q/g}=\e^{m'kl}\o_\pp^k\hat{q}_l A_2^{q/g}.
\end{align}
%\begin{flushleft}
%For fermion loop:
%\end{flushleft}
%\begin{align}
%A_1^f(q^0,q)=&-4i\frac{N_f}{2}(2\pi)^2\int_K\delta(K^2)\delta\left(\left(K+Q\right)^2\right)\varepsilon(k^0)\varepsilon(k^0+q^0)\frac{1}{q^2}\left[k_\parallel^2+qk_\parallel\right.\notag\\&-k^0(k^0+q^0)\Big]\left(1-\tilde{f}\left(k^0\right)\right)\tilde{f}^{\prime}(k^0+q^0)    
%\end{align}
%\begin{equation}
%A_2^f(q^0,q)=4i\frac{N_f}{2}(2\pi)^2\int_K\delta(K^2)\delta\left(\left(K+Q\right)^2\right)\varepsilon(k^0)\varepsilon(k^0+q^0)\frac{1}{q}k_\parallel(k^0+q^0)\left(1-\tilde{f}\left (k^0\right)\right)\tilde{f}^{\prime}(k^0+q^0)
%\end{equation}

%\begin{flushleft}
%For gluon loop:
%\end{flushleft}
%\begin{align}
%A_1^g(q^{0},q)=&-iN_C(2\pi)^{2}\int_{K}\delta(K^{2})\delta\left(\left(K+Q\right)^{2}\right)\varepsilon(k^{0})\varepsilon(k^{0}+q^{0})\left[\frac{k_{\perp}^{2}}{2k_0^{2}}\left(-8\frac{k_\parallel}{q}+4\frac{k_\parallel^2}{(k_0+q_0)^2}\right) \right.\notag \\
%&-\frac{k_\perp^2}{2}\left(-2+2\frac{k_\parallel+q}{(k_0+q_0)^2}q-4\frac{q^2}{(k_0+q_0)^2}\right)\Bigg]  f^{\prime}(k^{0})f\left(k^{0}+q_0\right)
%\end{align}
%\begin{align}
%A_2^g(q^0,q)=&-iN_C(2\pi)^2\int_K\delta(K^2)\delta\left(\left(K+Q\right)^2\right)\varepsilon(k^0)\varepsilon(k^0+q^0)2(2k^0+q^0)\left[\frac{k_\perp^2}{4k_0^2}(2-\frac{k_\parallel+q}{(k_0+q_0)^2}q)\right.\notag\\&+\frac{1}{2}(2+\frac{k_\parallel}{2q}-\frac{k_\perp^2}{4(k^0+q^0)^2})-\frac{(k_\parallel+q)^2}{(k^0+q^0)^2}\Bigg]f^{\prime}(k^{0})f\left(k^{0}+q_0\right)
%\end{align}
%
%\begin{flushleft}
%Here  $k_\parallel=\frac{q_{0}^{2}-q^{2}+2k_0q_0}{2q}$ , $k_\perp^2=k^2-k_\parallel^2$.
\begin{figure}
\includegraphics[width=0.45\linewidth]{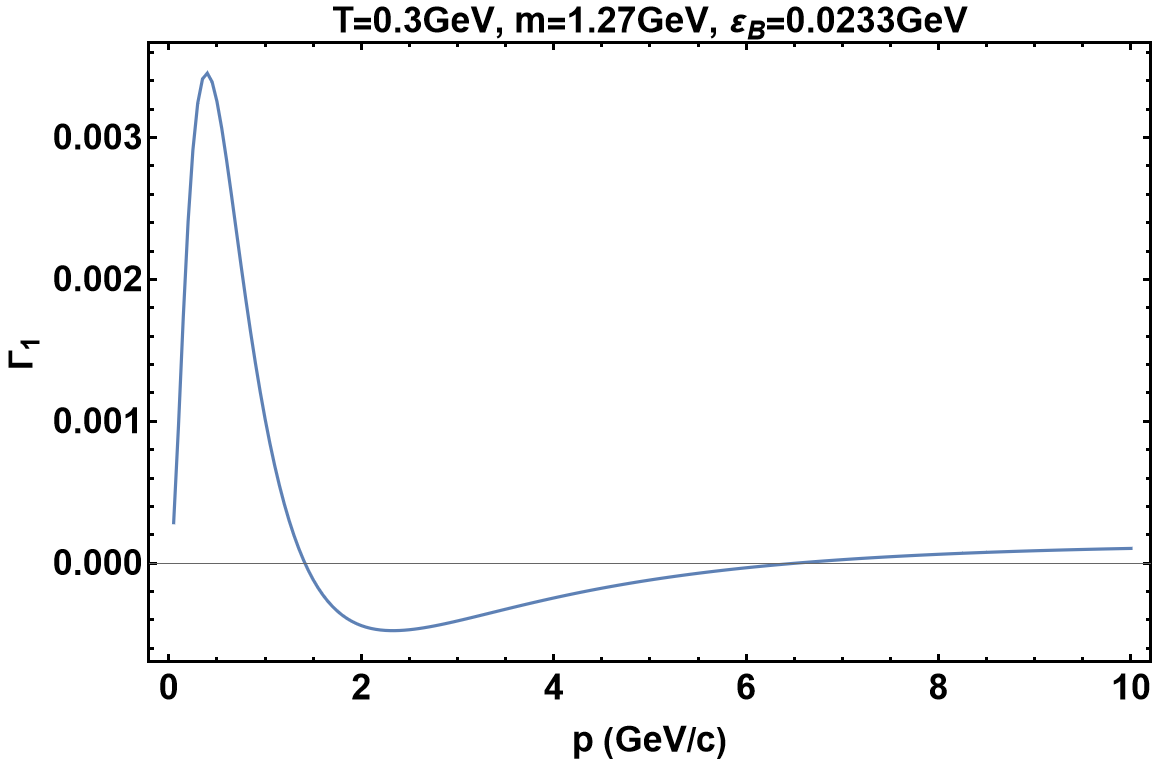}
\includegraphics[width=0.45\linewidth]{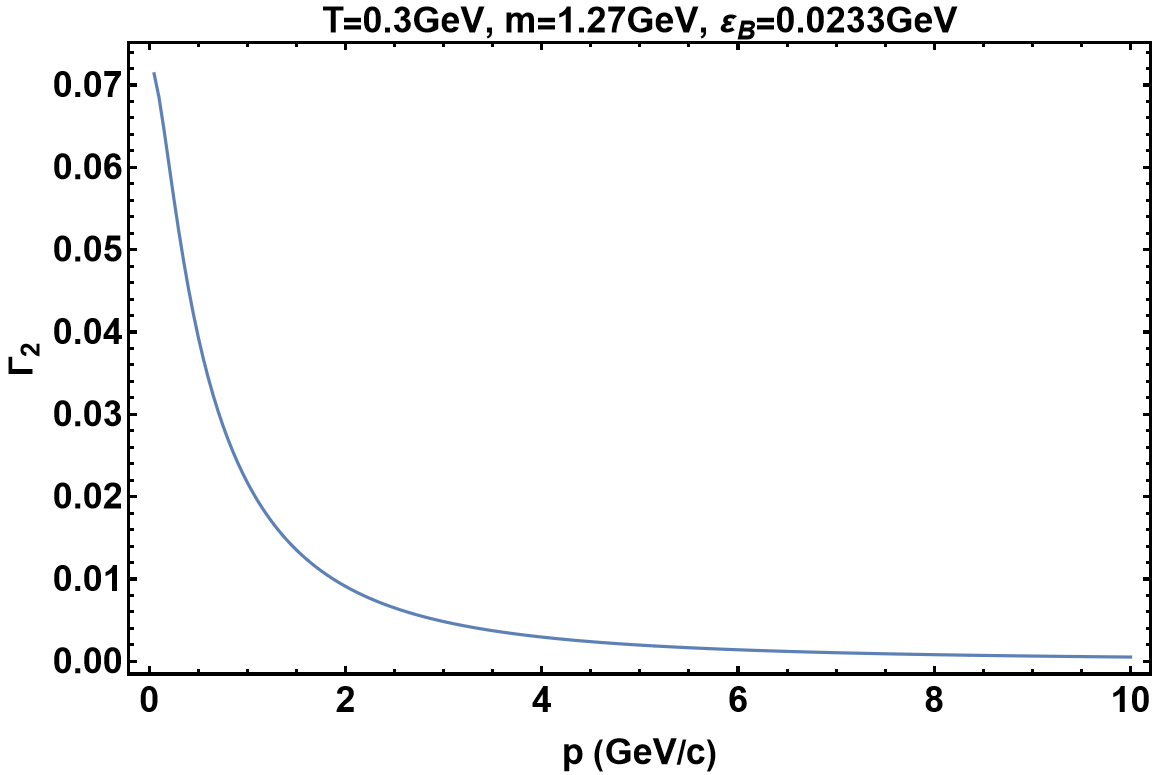}
\includegraphics[width=0.45\linewidth]{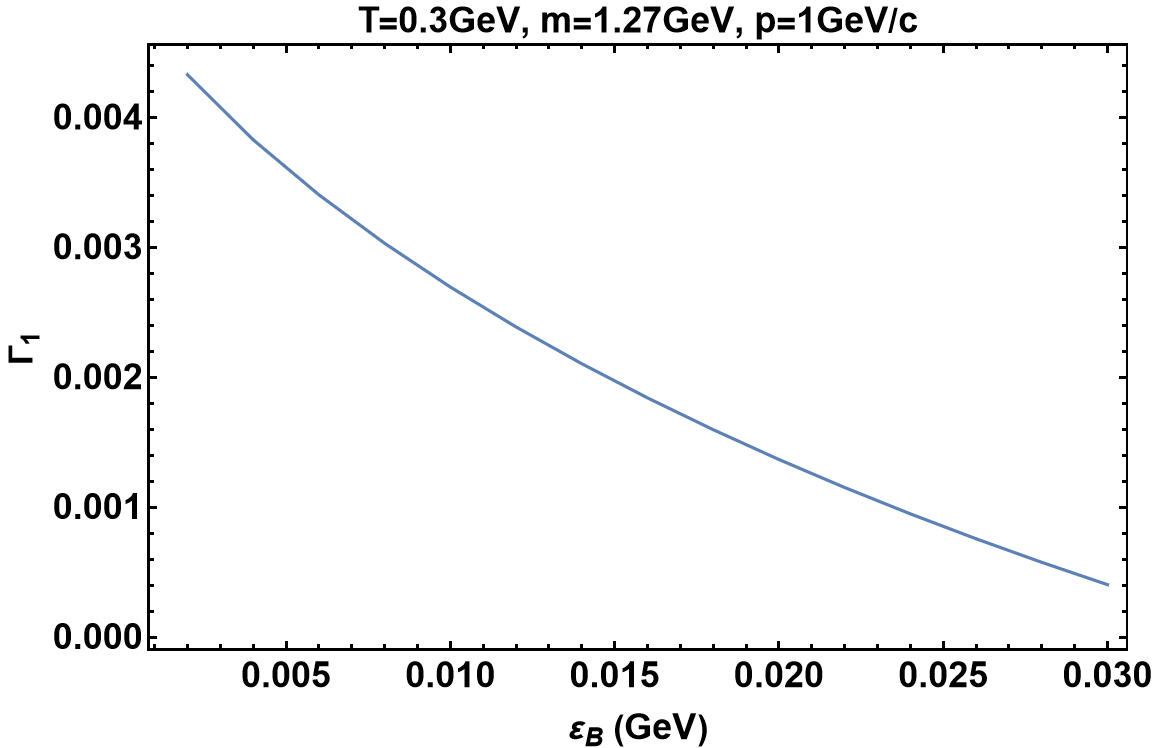}
\includegraphics[width=0.45\linewidth]{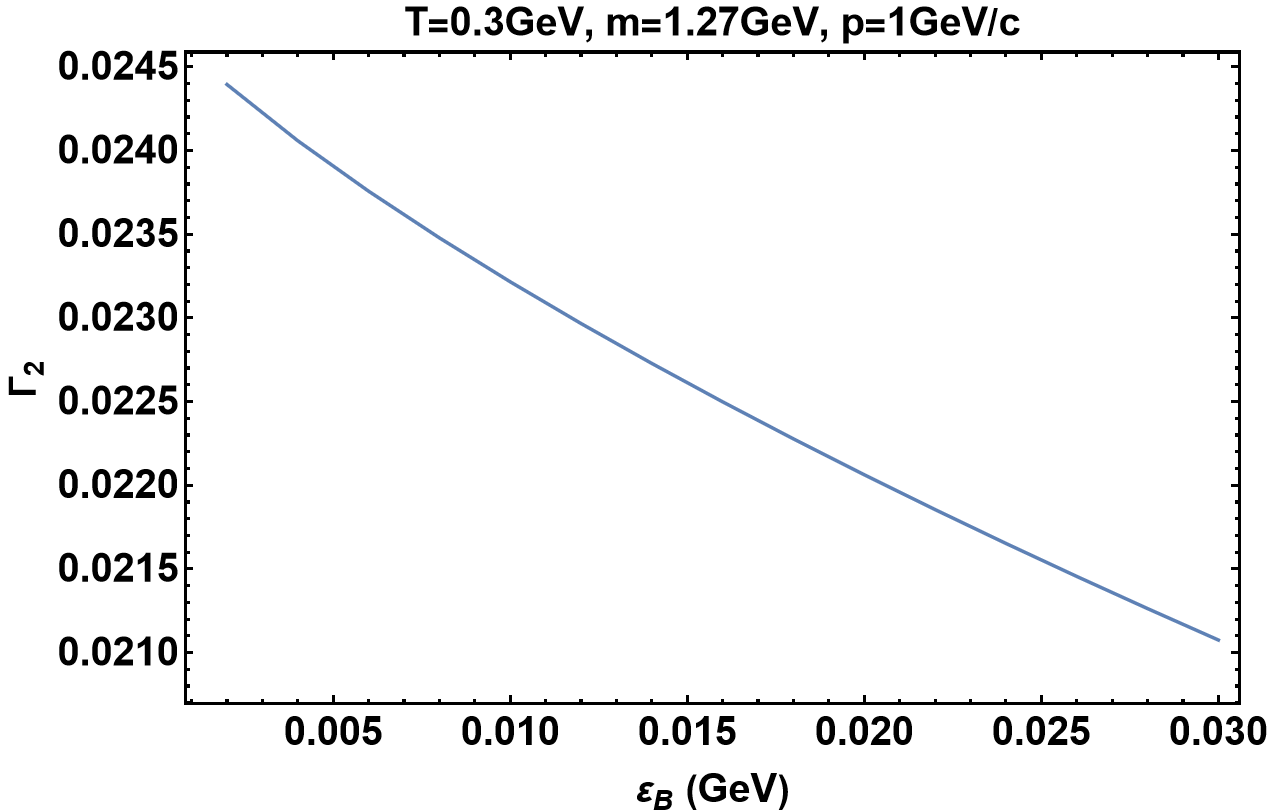}
\includegraphics[width=0.45\linewidth]{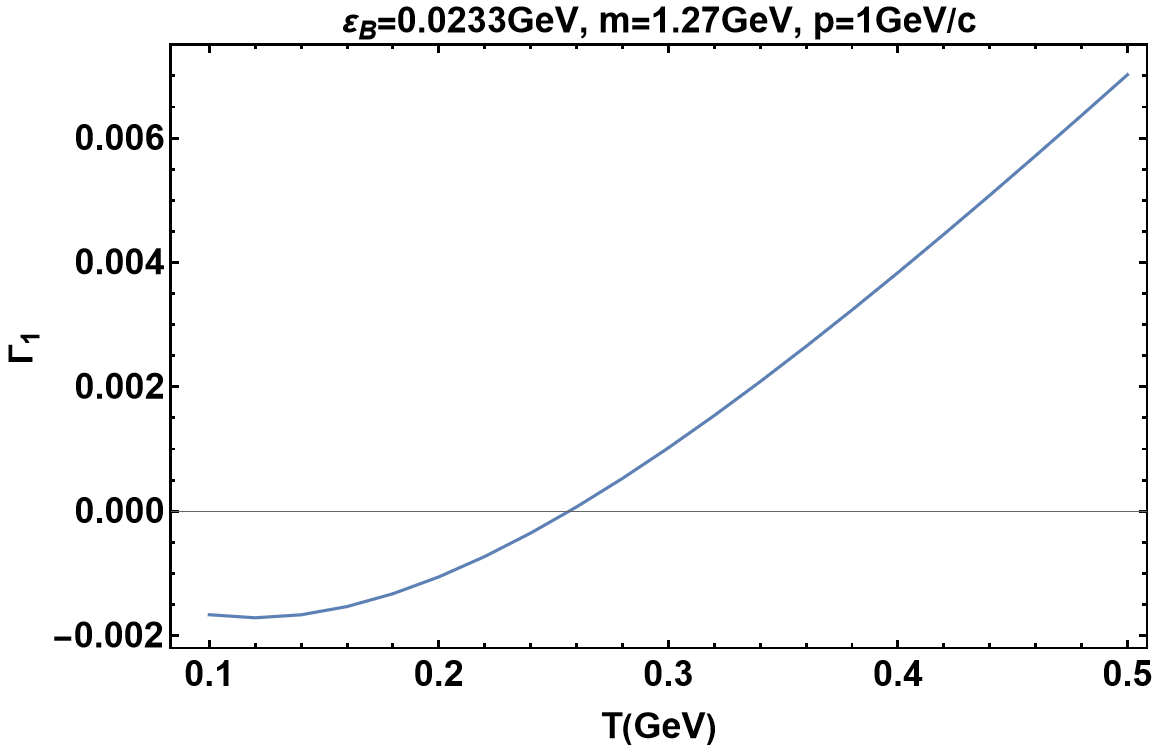}
\includegraphics[width=0.45\linewidth]{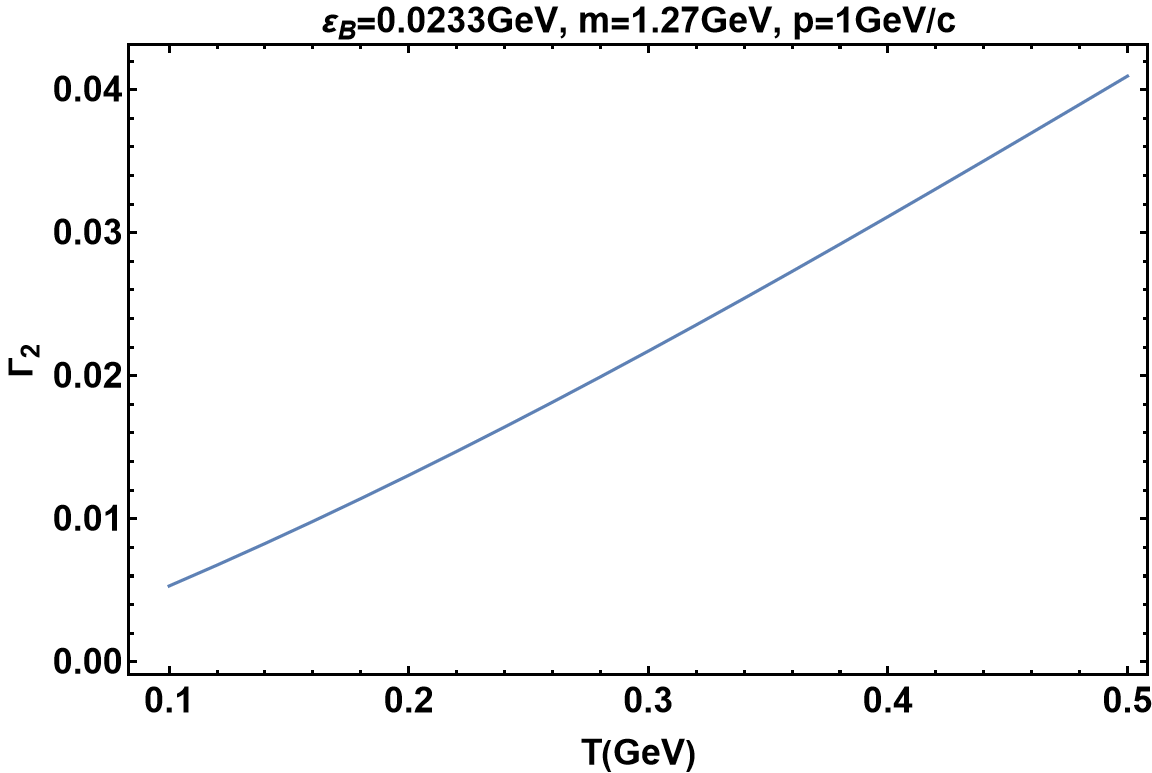}
    \caption{$\Gamma_1$ and $\Gamma_2$ defined in \eqref{Gamma12} as functions of $p$, $\e_b$ and $T$ respectively, with $\a_s=0.3$, $N_f=N_c=3$ and $m$ taken to be charm quark mass. $\e_b$ is temperature dependent. For simplicity, we take $\e_b=0.0233\,\text{GeV}$ for an average temperature of $210\,\text{MeV}$ corresponding to the weak binding scenario \cite{Zhao:2010nk}. We find $|\G_2|\gg|\G_1|$ numerically, indicating the angular dependence of $\G_s^{(1)}$ is largely set by $\o\cdot\bm{n}$, with only a small correction from dependence on quark momentum direction $\hat{p}$.}
    \label{fig:Gammas}
\end{figure}
In Fig.~\ref{fig:Gammas}, we show $\G_{1}$ and $\G_2$ as functions of $p$, $\e_b$ and $T$ with $\a_s=0.3$, $N_f=N_c=3$ and $m$ taken to be charm quark mass. For simplicity, we take $\e_b=0.0233\,\text{GeV}$ for an average temperature of $210\,\text{MeV}$ corresponding to the weak binding scenario \cite{Zhao:2010nk}. We find $|\G_2|\gg|\G_1|$, indicating the angular dependence of $\G_s^{(1)}$ is largely set by $\o\cdot\bm{n}$, with only a small correction from dependence on quark momentum direction $\hat{p}$. $\G_1$ tends to zero as $p\to 0$, which is consistent with the expectation of no $\hat{p}$-dependence in this limit. %$\G_{1,2}$ show milder dependencies on $T$ in the phenomenologically relevant temperature range of QGP.

A surprising result is the absence of $\ln\frac{T}{\e_b}$ enhancement, which we now explain. Note that $\ln\frac{T}{\e_b}$ enhancement has indeed been found in earlier calculations of spin independent dissociation rate, see \cite{Brambilla:2010vq}. It arises from quark coupling to fluctuations of incompletely screened chromomagnetic fields. Since there are two spin-magnetic vertices, the dependence on spin cancels in the product. In our case, the fluctuations of chromomagnetic field can be identified with terms $\propto|\D_T|^2$ in \eqref{Gamma_1} and \eqref{Gamma_2}. In the mean time, we have calculated explicitly the IR limit of vortical correction to gluon self-energy in appendix~\ref{sec_app_C}, finding the same scaling with $q$ as their counterpart in the absence of vorticity. However, a similar $\ln\frac{T}{\e_b}$ does not occur in case of vortical correction due to presence of an additional factor of $q_0$ in the corresponding terms. The appearance of $q_0$ can be understood as follows: we are after a spin dependent dissociation rate. The quark can only have spin dependent coupling to chromomagentic field in one vertex and spin independent coupling to chromoelectric field in the other vertex\footnote{See \cite{Bodwin:1994jh} for explicit form of couplings in the non-relativistic expansion.}, so that the product is still spin dependent. The factor of $q_0$ is necessary for converting one of the transverse gauge field in $D^{m'n'<(1)}=-|\D_T|^2P_T^{m'm}P_T^{n'n}\P_{q/g}^{mn<(1)}$ into a chromoelectric field. Since chromoelectric field is completely screened, no logarithmic enhancement is found.

So far, we have only calculated dissociation from scattering with one of the constituent quark. In the quasi-free picture, we can easily obtain the dissociation rate for quarkonia by adding contributions from two constituents
\begin{align}
    \G^{(1)}_+=2\G^{(1)}_{1/2},\quad \G^{(1)}_-=2\G^{(1)}_{-1/2},\quad \G^{(1)}_0=\G^{(1)}_{1/2}+\G^{(1)}_{-1/2}=0.
\end{align}

\section{Spin alignment of $\jp$ in vortical QGP}\label{sec_pheno}

\begin{figure}
    \centering
    \includegraphics[width=0.5\linewidth]{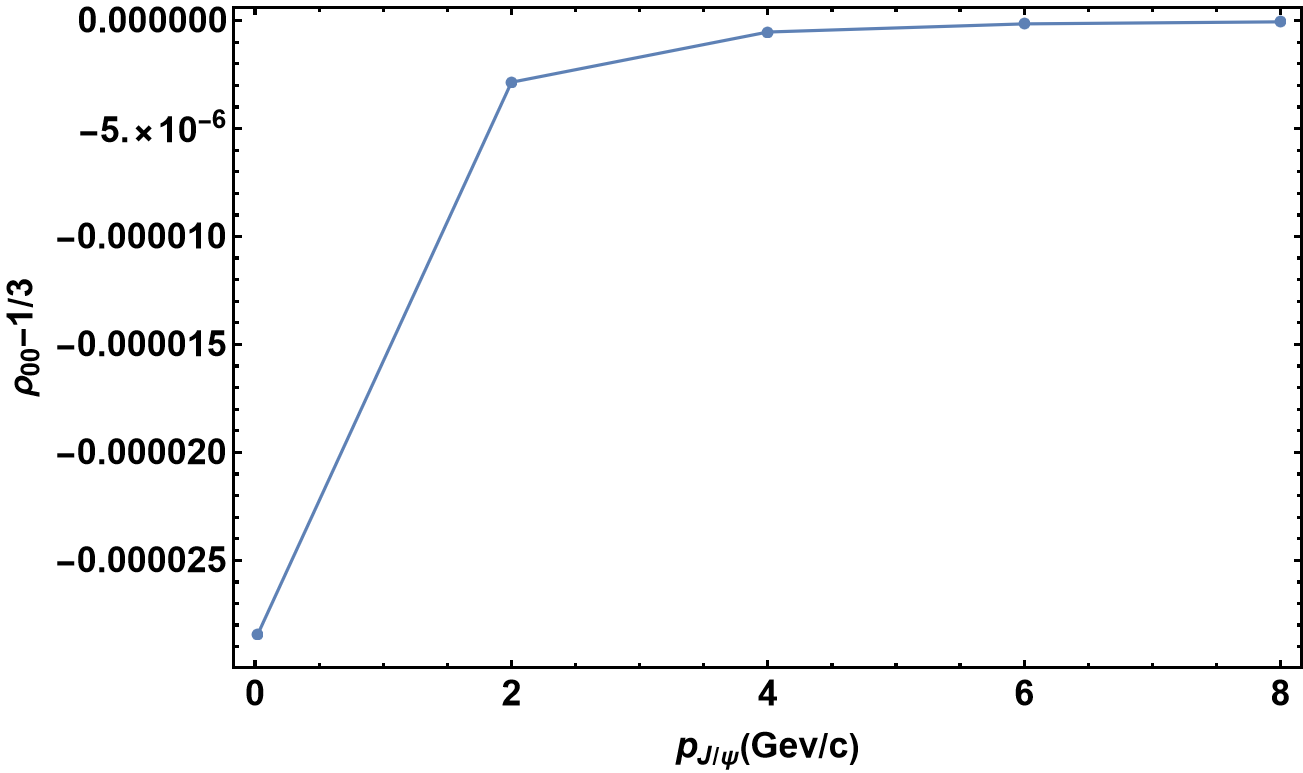}
    \caption{Spin alignment $\rho_{00}-1/3$ as a function of $p$.}
    \label{fig:alignment}
\end{figure}

To implement the spin-dependent dissociation rate in spin alignment for $\jp$, we take Bjorken flow for the evolution of QGP and take the evolution of vorticity from Eq.(8) of \cite{Jiang:2016woz}, see also \cite{Deng:2016gyh}
\begin{align}
&\o(t,b,\rts)=\tanh (0.28 b) (0.001775 \tanh (3-0.015 \rts)+0.0128) (\exp (-0.016 b \rts)+1)+\nonumber\\
&(0.02388 b+0.01203) (0.58 t)^{0.35} \exp (-0.58 t) (1.751\, -\tanh (0.01 \rts)) (\exp (-0.016 b \rts)+1)
\end{align}
For illustration purpose, we consider quarkonia spin alignment in collisions with $\rts=200\,\text{GeV}$. The temperature profile is modeled by Bjorken flow with
\begin{align}
    T=T_0\(\frac{\t}{\t_0}\)^{-1/3},
\end{align}
with $T_0=350\,\text{MeV}$ and $\t_0=0.6\,\text{fm}$. $\t=\sqrt{t^2-z^2}$ is the proper time. Given that $|\G_1|\ll|\G_2|$ numerically, we shall ignore the term dependent on quark momentum direction $\hat{p}$. Since the quark and anti-quark are supposed to carry close momenta in the quasi-free picture, this means the spin dependent dissociation rate is nearly independent of quarkonia momentum direction. We can then obtain a simple form of spin alignment as follows: assuming a spin independent distribution function $f$ for initially produced quarkonia and considering dissociation contribution only, we find the distribution of quarkonia in the $i$-th spin state given by
\begin{align}
    f_i=e^{-\int (\G_0+\G_i^{(1)})d\t}f,
\end{align}
with $i=0,+,-$. Further assuming $\bm{n}\pr \o$, we find the corresponding spin alignment given by
\begin{align}
    &\r_{00}-\frac{1}{3}=\frac{f_0}{f_0+f_++f_-}-\frac{1}{3}\nonumber\\
    =&\frac{1}{1+e^{\int \G_2d\t}+e^{-\int \G_2\t}}-\frac{1}{3}\simeq-\frac{1}{9}\(\int\G_2d\t\)^2.
\end{align}
In the last step, we have expanded to first non-trivial order in $\int \G_sd\t$. The $\t$-integration is chosen to start at $\t_0$ and stop at $T=150\,\text{MeV}$. We show in Fig.~\ref{fig:alignment} the $p$-dependence of spin alignment.

\section{Conclusion and outlook}\label{sec_concl}

We have considered dissociation of quarkonia in a vortical QGP and proposed spin dependent dissociation rate as a possible mechanism for quarkonia spin alignment observed in heavy ion experiments. In this mechanism, small vorticity in high energy collisions may be compensated by long evolution time of QGP. With QGP constituents being polarized by vorticity, inelastic scattering between quarkonium constituents and QGP constituents naturally leads to a spin dependent dissociation rate for quarkonia. Aiming at leading logarithmically enhanced contribution in binding energy, we focus on Coulomb scattering process in this work. We have obtained non-trivial angular structure of the vortical correction to dissociation rate that depends on directions of quantization axis, vorticity and constituent quark momentum. The dependence on quark momentum direction is found to be weak for phenomenologically interesting parameter choices. We have implemented the results in a simplified dissociation dominated evolution of quarkonia, which gives a slightly suppressed spin zero states compared to the average of the other two states.

A surprising result we have found both numerically and analytically is the absence of the logrithmical enhancement in the binding energy in the vortical correction to the dissociation rate. This is understood as from the requirement that a spin dependent dissociation can only come from quark coupling to a pair of chromomagnetic and chromoelectric field. In this case, screening effect is sufficient to render the results free of logarithmical enhancement.

Clearly a complete study requires inclusion of Compton scattering. Indeed, early calculations of heavy quark energy loss \cite{Peigne:2008nd} indicates a possible significant contribution from Compton scattering, suggesting a similar contribution to the dissociation rate. A more interesting questions how do other hydrodynamic gradients affect the spin alignment. A specific example is the shear, which can sustain for long time and is known to contribute to local spin polarization significantly \cite{Fu:2021pok,Becattini:2021iol,Yi:2021ryh}. It would certainly be desirable to study spin dependent dissociation in a QGP with shear flow. We leave these interesting directions for future explorations.

%{\large {\bf Acknowledgment}}
\section*{Acknowledgments}
We thank X.-z. Bai, P. Braun-Munzinger, Y. Guo, Y. Jia, M. He, D.-f. Hou, K. Zhou, P.-f. Zhuang for stimulating discussions. This work is in part supported by NSFC under Grant Nos 12475148, 12075328.

\appendix
\section{Vortical correction to gluon self-energy}\label{sec_app_A}

We start with the evaluation of vortical correction to gluon self-energy from quark loop. Working in the QGP frame with $u^\m=(1,0,0,0)$, we can spell out explicit form of \eqref{Pi_quark} as
\begin{align}
&\P^{ij<(1)}_q=-2\fctrq\int_K\big[-4i\e^{ijk}(k+q)_l\o_lk_k+4i\e^{ijk}(k+q)_0k_0\o_k\big](2\p)^2\d((K+Q)^2)\d(K^2)\e(k_0)\e(k_0+q_0)\nonumber\\
&\times\widetilde{f}'(k_0+q_0)\(1-\widetilde{f}(k_0)\),\nonumber\\
&\P^{0i<(1)}_q=-2\fctrq\big[4i\e^{ijk}(k+q)_0\o_j k_k\big](2\p)^2\d((K+Q)^2)\d(K^2)\e(k_0)\e(k_0+q_0)\widetilde{f}'(k_0+q_0)\(1-\widetilde{f}(k_0)\).
\end{align}
Note that the heavy quarks are on-shell in the quasi-free picture $P^2=(P+Q)^2=m_Q^2$. The gluon energy threshold is set by the binding energy $q_0\geq \e_B>0$. We can easily show that the gluon is spacelike as long as $Q\ll m_Q$. Using $\d(K^2)$, we have $\d((K+Q)^2)=\d(q_0^2-q^2+2k_0q_0-2k q\cos\th)$, with $\th$ being the angle between $\bm{k}$ and $\bm{q}$. $\d((K+Q)^2)$ fixes $\cos\th=\frac{q_0^2-q^2+2k_0q_0}{2kq}$. $|\cos\th|\leq 1$ gives the following kinematic constraint on $k$
\begin{align}
2k\geq q-q_0\quad \text{for}\; k_0=k;\quad 2k\geq q+q_0\quad \text{for}\; k_0=-k.
\end{align} 
For both cases, we have $\e(k_0)\e(k_0+q_0)=1$. To simplify the tensor structures, we note that $\P^{ij<(1)}$ is to be contracted with transverse projectors on both sides, so we may simply take $i$ and $j$ to be transverse indices. It follows that the index $k$ has to be longitudinal. Formally, we can express this statement by the following identity
\begin{align}\label{projection_identity}
\e^{ijk}P^{im}_TP^{jn}_TP^{kl}_T=0\;\Rightarrow\;
\e^{ijk}P^{im}_TP^{jn}_T\d^{kl}=\e^{ijk}P^{im}_TP^{jn}_T\hat{q}^k\hat{q}^k.
\end{align}
The first identity can be shown by expanding it out and using the Schouten identity $\e^{inl}\hat{q}^m-\e^{nlm}\hat{q}^i+\e^{lmi}\hat{q}^n-\e^{min}\hat{q}^l=0$. We can then obtain the following projected $\P^{ij<(1)}$ as
\begin{align}\label{project_quark1}
&P^{m'm}_TP^{n'n}_T\P^{mn<(1)}_q=-2\fctrq\int_K\big[-4i\e^{m'n'k}(k+q)_l\o_l k_k^\pr+4i\e^{m'n'k}(k+q)_0k_0\o_k^\pr\big](2\p)^2\d((K+Q)^2)\d(K^2)\nonumber\\
&\times\widetilde{f}'(k_0+q_0)\(1-\widetilde{f}(k_0)\)\\
&=-2\fctrq\int_K\big[-4i\e^{m'n'k}\hat{q}_k\hat{q}_l\o_l\((k^\pr+q) k^\pr-(k+q)_0k_0\)\big](2\p)^2\d((K+Q)^2)\d(K^2)\widetilde{f}'(k_0+q_0)\(1-\widetilde{f}(k_0)\)\nonumber.
\end{align}
We use $\pr$ to denote longitudinal projection. $k^\pr=k\cdot\hat{q}$. In the second line, we have made the replacement $(k+q)_l\to(k^\pr+q)_l$. This is because $k^\pp_\l k^\pr_k$ vanishes up on angular integration of $\bm{k}$.
Similarly $\P^{0i<(1)}_q$ is projected as
\begin{align}\label{project_quark2}
P_T^{m'm}\P^{0m<(1)}_q=-2\fctrq\int_K\big[4i\e^{m'jk}(k+q)_0\o_j\hat{q}_k k_\pr\big](2\p)^2\d((K+Q)^2)\d(K^2)\widetilde{f}'(k_0+q_0)\(1-\widetilde{f}(k_0)\).
\end{align}

Now we turn to vortical correction from gluon loop. Noting that vortical correction to either gluon propagator gives identical contribution and gluon propagators in the loop take only spatial indices, we may write
\begin{align}
&\Pi^{\alpha\beta<(1)}_g=-2\fctrg\frac{1}{2}\int_K(2\p)^2\d((K+Q)^2)\d(K^2)\e(k_0)\e(k_0+q_0)P_{ij}^T(K+Q)\big[\frac{i\e^{kla}k_ak_b\o_b}{2k_0^2}+\frac{i\e^{kla}\o_a}{2}\big]f'\nonumber\\
&\Big[g^{i\alpha}(-K-2Q)^l+g^{\alpha l}(Q-K)^i+g^{li}(2K+Q)^\alpha\Big]\Big[g^{j\beta}(K+2Q)^k+g^{\beta k}(K-Q)^j+g^{kj}(-2K-Q)^\beta\Big].
\end{align}
The product of vertices in the second line above can be written out explicitly. For $\a\b=0m$, we have
\begin{align}\label{product_0m}
&\d^{il}\d^{jm}(2k+q)_0(k+2q)_k+\d^{il}\d^{mk}(2k+q)_0(-2q_j)+\d^{il}\d^{kj}(2k+q)_0(-2k-q)_m\nonumber\\
=&\d^{il}\d^{jm}(2k+q)_0(k+2q)_k+\d^{il}\d^{mk}(2k+q)_0(-2q_j),
\end{align}
where the last term is dropped up on contraction with $P_{ij}^T(K+Q)\big[\frac{i\e^{kla}k_ak_b\o_b}{2k_0^2}+\frac{i\e^{kla}\o_a}{2}\big]$, which is symmetric in $ij$ and anti-symmetric in $kl$. For $\a\b=mn$, the product is a little complicated:
\begin{align}\label{product_mn}
&\d^{im}\d^{jn}(-k-2q)_l(k+2q)_k+\d^{im}\d^{nk}(-k-2q)_l(-2q_j)+\d^{im}\d^{kj}(-k-2q)_l(-2k-q)_n\nonumber\\
+&\d^{ml}\d^{jn}2q_i(k+2q)_k+\d^{ml}\d^{nk}2q_i(-2q_j)+\d^{ml}\d^{kj}2q_i(-2k-q)_n\nonumber\\
+&\d^{il}\d^{jn}(2k+q)_m(k+2q)_k+\d^{il}\d^{nk}(2k+q)_m(-2q_j)+\d^{il}\d^{kj}(2k+q)_m(-2k-q)_n.
\end{align}
The first and last terms are symmetric in $mn$ thus can be dropped\footnote{The anti-symmetric part of the contribution from $D_{ij}^{<(0)}(K+Q)D_{kl}^{>(1)}(K)$ is doubled by the counterpart from $D_{ij}^{<(1)}(K+Q)D_{kl}^{>(0)}(K)$, while the symmetric part is canceled. This can be seen easily from the discussion below \eqref{Pi_gluon}.}. The remaining terms can be organized as
\begin{align}
&\d^{im}\d^{nk}(-k-2q)_l(-2q_j)+\d^{im}\d^{kj}(-k-2q)_l(-2k-q)_n+\d^{ml}\d^{kj}2q_i(-2k-q)_n-(m\leftrightarrow n)\nonumber\\
&+\d^{ml}\d^{nk}2q_i(-2q_j).
\end{align}
We then contract \eqref{product_0m} and \eqref{product_mn} with $P_{ij}^T(K+Q)\big[\frac{i\e^{kla}k_ak_b\o_b}{2k_0^2}+\frac{i\e^{kla}\o_a}{2}\big]$ and project the spatial indices to the transverse plane. For $\a\b=0m$, we have
\begin{align}\label{contraction_0m}
&\P_T^{m'm}(Q)\d^{il}\d^{jm}(2k+q)_0(k+2q)_k\big[\d_{ij}-\frac{(k+q)_i(k+q)_j}{(\bm{k}+\bm{q})^2}\big]\big[\frac{i\e^{kla}k_ak_b\o_b}{2k_0^2}+\frac{i\e^{kla}\o_a}{2}\big]\nonumber\\
=&(2k+q)_02q_k\frac{i\e^{km'a}k_\pp^2\o_a^\pp}{4k_0^2}+(2k+q)_0(k_\pr+2q)_k\frac{i\e^{km'a\o_a^\pp}}{2}+(2k+q)_0\frac{k_\pp^2}{2}\frac{q_l}{(k_0+q_0)^2}\frac{i\e^{m'la}\o_a^\pp}{2},\nonumber\\
&\P_T^{m'm}(Q)\d^{il}\d^{mk}(2k+q)_0(-2q_j)\big[\d_{ij}-\frac{(k+q)_i(k+q)_j}{(\bm{k}+\bm{q})^2}\big]\big[\frac{i\e^{kla}k_ak_b\o_b}{2k_0^2}+\frac{i\e^{kla}\o_a}{2}\big]\nonumber\\
=&(2k+q)_0(-2q_l)\(1-\frac{(k_\pr+q)q}{(k_0+q_0)^2}\)\frac{i\e^{m'la}k_\pp^2\o^\pp_a}{4k_0^2}+(2k+q)_0(-2q_l)\(1-\frac{(k_\pr+q)^2}{(k_0+q_0)^2}\)\frac{i\e^{m'la}\o_a^\pp}{2},
\end{align}
where $\o_a^\pp=\(\d_{ab}-\hat{q}_{a}\hat{q}_{b}\)\o_b$. In arriving at the above, we have used $\d((K+Q)^2)$ and made replacement like $k_a^\perp k_b^\perp\to\frac{1}{2}k_\pp^2\(\d_{ab}-\hat{q}_{a}\hat{q}_{b}\)$ by rotational invariance in the transverse plane. Similarly for $\a\b=mn$, we have
\begin{align}\label{contraction_mn}
&P_T^{m'm}P_T^{n'n}\d^{im}\d^{nk}(-k-2q)_l(-2q_j)\big[\d_{ij}-\frac{(k+q)_i(k+q)_j}{(\bm{k}+\bm{q})^2}\big]\big[\frac{i\e^{kla}k_ak_b\o_b}{2k_0^2}+\frac{i\e^{kla}\o_a}{2}\big]\nonumber\\
=&2\frac{(k_\pr+q)q}{(k_0+q_0)^2}\big[\frac{i\e^{n'lm'}k_\pp^2k_\pr\o_l^\pr}{4k_0^2}(-2q)+\frac{i\e^{n'm'a}\o_a^\pr}{2}\(-\frac{1}{2}k_\pp^2\)\big],\nonumber\\
&P_T^{m'm}P_T^{n'n}\d^{im}\d^{kj}(-k-2q)_l(-2k-q)_n\big[\d_{ij}-\frac{(k+q)_i(k+q)_j}{(\bm{k}+\bm{q})^2}\big]\big[\frac{i\e^{kla}k_ak_b\o_b}{2k_0^2}+\frac{i\e^{kla}\o_a}{2}\big]\nonumber\\
=&\frac{i\e^{m'ln'}k_\pr\o_l^\pr}{2k_0^2}2q k_\pp^2+\frac{i\e^{n'm'a}\o_a^\pr}{2}k_\pp^2,\nonumber\\
&P_T^{m'm}P_T^{n'n}\d^{ml}\d^{kj}2q_i(-2k-q)_n\big[\d_{ij}-\frac{(k+q)_i(k+q)_j}{(\bm{k}+\bm{q})^2}\big]\big[\frac{i\e^{kla}k_ak_b\o_b}{2k_0^2}+\frac{i\e^{kla}\o_a}{2}\big]\nonumber\\
=&2q\frac{k_\pp^2k_\pr}{(k_0+q_0)^2}(-k_\pp^2)\frac{i\e^{km'n'}\o_k^\pr}{2k_0^2}+\frac{2q(k_\pr+q)}{(k_0+q_0)^2}k_\pp^2\big[\frac{i\e^{n'm'a}\o_a^\pr k_\pr^2}{2k_0^2}+\frac{i\e^{n'm'a}\o_a^\pr}{2}\big],\nonumber\\
&P_T^{m'm}P_T^{n'n}\d^{ml}\d^{nk}2q_i(-2q_j)\big[\d_{ij}-\frac{(k+q)_i(k+q)_j}{(\bm{k}+\bm{q})^2}\big]\big[\frac{i\e^{kla}k_ak_b\o_b}{2k_0^2}+\frac{i\e^{kla}\o_a}{2}\big]\nonumber\\
=&-4\frac{q^2k_\pp^2}{(k_0+q_0)^2}\big[\frac{i\e^{n'm'a}\o_a^\pr k_\pr^2}{2k_0^2}+\frac{i\e^{n'm'a}\o_a^\pr}{2}\big].
\end{align}
Collecting all terms, we have
\begin{align}\label{project_gluon}
&P_T^{m'm}\P^{0m<(1)}_g=-\fctrg\int_K(2k+q)_02q_l\Bigg[\(2-\frac{(k_\pr+q)q}{(k_0+q_0)^2}\)\frac{i\e^{lm'a}k_\pp^2\o_a^\pp}{4k_0^2}+\(2+\frac{k_\pr}{2q}-\frac{k_\pp^2+2(k_\pr+q)^2}{4(k_0+q_0)^2}\)\nonumber\\
&\times\frac{i\e^{lm'a}\o_a^\pp}{2}\Bigg](2\p)^2\d((K+Q)^2)\d(K^2)f(k_0+q_0)f'(k_0),\nonumber\\
&P_T^{m'm}P_T^{n'n}\P^{mn<(1)}_g=-\fctrg\int_K\Bigg[i\e^{n'lm'}\o_l^\pr\(\frac{k_\pp^2k_\pr q}{2k_0^2}+\frac{k_\pp^2}{4}\)\(-4\frac{(k_\pr+q)q}{(k_0+q_0)^2}-4\)+4q\frac{k_\pp^2k_\pr}{(k_0+q_0)^2}(-k_\pp^2)\nonumber\\
&\times\frac{i\e^{km'n'}\o_k^\pr}{2k_0^2}+\frac{4q k_\pr k_\pp^2}{(k_0+q_0)^2}\(\frac{i\e^{n'm'a}\o_a^\pr k_\pr^2}{2k_0^2}+\frac{i\e^{n'm'a}\o_a^\pr}{2}\)\Bigg](2\p)^2\d((K+Q)^2)\d(K^2)f(k_0+q_0)f'(k_0).
\end{align}

\section{Angular average of $q$}\label{sec_app_B}
We start by writing out explicit expressions of \eqref{trace} as
\begin{align}
    &\tr[u_+(P)\bar{u}_+(P)\g^\m({\slashed P}+{\slashed Q})\g^\n D_{\n\m}^{<(1)}(Q)]\nonumber\\
&=-(p_0+m)\big[i\e^{nmk}(p+q)_k D_{mn}^{<(1)}(Q)2\pmh\cdot\hat{n}+(i\e^{nmi}(p+q)_0D_{mn}^{<(1)}+2i\e^{mki}(p+q)_kD_{0m}^{<(1)})\times\nonumber\\
&\(-n_i(1-\pmh^2)+2\pmh\cdot\hat{n}\pmh_i\)\big],\nonumber\\
&\tr[u_+(P)\bar{u}_+(P)(-i\S^{\m\n})D_{\n\m}^{<(1)}(Q)m]\nonumber\\
&=-(p_0+m)\big[-2iD_{i0}^{<(1)}(Q)2m\e^{ijk}\pmh_j n_k+iD_{ji}^{<(1)}(Q)m\e^{ijk}\(n_k(1+\pmh^2)-2\pmh\cdot\hat{n}\pmh_k\)\big].
\end{align}
Using \eqref{A12_def}, we may further simplify the above as
\begin{align}
  &\tr[u_+(P)\bar{u}_+(P)\g^\m({\slashed P}+{\slashed Q})\g^\n D_{\n\m}^{<(1)}(Q)]\nonumber\\
&=(p_0+m)\Big[iA_1|\D_T|^2\(2(-p-q)_k\o_k^\pr2\pmh\cdot\hat{n}+2(p+q)_0\o_i^\pr\(n_i(1-\pmh^2)+2\pmh\cdot\hat{n}\pmh_i\)\)\nonumber\\
&+iA_2\frac{\D_T\D_T^*+\D_T^*\D_L}{2}2(-p-q)_k\(\hat{q}_k\o_i^\pp-\hat{q}_i\o_k^\pp\)\(n_i(1-\pmh^2)+2\pmh\cdot\hat{n}\pmh_i\)\Big],\nonumber\\
&\tr[u_+(P)\bar{u}_+(P)(-i\S^{\m\n})D_{\n\m}^{<(1)}(Q)m]\nonumber\\
&=(p_0+m)\Big[iA_2\frac{\D_T\D_T^*+\D_T^*\D_L}{2}4m(\hat{q}_j\o_k^\pp-\hat{q}_k\o_j^\pp\pmh_j\hat{n}_k)\nonumber\\
&-iA_1|\D_T|^2 2m\o_k^\pr\(n_k(1-\pmh^2)+2\pmh\cdot\hat{n}\pmh_k\)\Big].
\end{align}
To perform angular integration of $q$, we decompose $q_k=q_k^\pr+q_k^\pp$ with the parallel and perpendicular components defined with respect to $\hat{p}$. Note that $q_\pr=\frac{q_0^2-q^2+2p_0q_0}{2p}$ is already fixed by $\d((P+Q)^2-m^2)=\d(2P\cdot Q+Q^2)$. It follows that we only need to perform angular integration in the perpendicular plane. Recall that $\o^\pr$ and $\o^\pp$ are defined with respect to $\hat{q}$ instead. The angular integration can be easily performed following the same logic as angular integration of $k$ in appendix~\ref{sec_app_A}, i.e. keeping only even power of $q_k^\pp$ and replace $q_i^\pp q_j^\pp\to\frac{1}{2}q_\pp^2\d_{ij}^\pp$ with $\d_{ij}^\pp=\d_{ij}-\hat{p}_i\hat{p}_j$. We obtain after straightforward calculations
\begin{align}
    &\pmh_i\o_i^\pr\to\pmh\cdot\o\cos\varphi,\nonumber\\
&(p+q)_i\o_i^\pr\to(p\cos\varphi+q)\hat{p}\cdot\o\cos\varphi,\nonumber\\
    &\o_i^\pr n_i\to\frac{1}{2}\big[\o\cdot n\sin^2\varphi+\(3\cos^2\varphi-1\)\hat{p}\cdot\o\,\hat{p}\cdot\hat{n}\big],\nonumber\\
    &\o_i^\pp n_i\to\(1-\frac{1}{2}\sin^2\varphi\)\o\cdot\hat{n}+\(\frac{1}{2}\sin^2\varphi-\cos^2\varphi\)\hat{p}\cdot\o\,\hat{p}\cdot\hat{n},\nonumber\\
    &\o_i^\pp \pmh_i\to\pmh\cdot\o\sin^2\varphi,\nonumber\\
    &p_i\o_i^\pp\hat{q}\cdot\hat{n}\to-\frac{1}{2}p\cos\varphi\sin^2\varphi\o\cdot\hat{n}+\frac{3}{2}\hat{p}\cdot\o\,\hat{p}\cdot\hat{n}\cos\varphi\sin^2\varphi,
\end{align}
with $\cos\varphi=\hat{q}\cdot\hat{p}$.

\section{IR limit of vortical correction to gluon self-energy}\label{sec_app_C}

We will need the IR limit of \eqref{project_quark1}, \eqref{project_quark2} and \eqref{project_gluon}. For the quark loop contribution, the dominant contribution comes from the phase space with $K\sim T\gg Q$, which leads to the following approximation
\begin{align}\label{Pi_quark_approx}
&P_T^{m'm}(Q)\P^{0m<(1)}_q\simeq-2\fctrq\int_0^\infty k^2dk\frac{1}{2\p}\(4i\e^{m'jk}k^2\o_j^\pp\hat{q}_k\frac{q_0}{q}\)\frac{1}{2kq}\frac{1}{2k}\widetilde{f}'(k),\nonumber\\
&P_T^{m'm}(Q)P_T^{n'n}\P^{mn<(1)}_q\simeq-2\fctrq\int_0^\infty k^2dk\frac{1}{2\p}(4i\e^{m'n'k}\o_k^\pr k^2\(1-\frac{q_0^2}{q^2}\))\frac{1}{2kq}\frac{1}{2k}\widetilde{f}'(k),
\end{align}
with the subscript ``q'' denoting quark loop contribution. In the above, the angular part of the $k$-integration is calculated as
\begin{align}
\int d\O\d((K+Q)^2)=\int d\cos\th d\f\d(q_0^2-q^2+2k_0q_0-2kq\cos\th)=2\p\frac{1}{2kq}.
\end{align}
The two pole contributions of $\d(K^2)$ is combined using $\widetilde{f}(k_0)+\widetilde{f}(-k_0)=-1$.
\eqref{Pi_quark_approx} scales as $T^2\o/q$ with correction suppressed by $O(q/T)$. A similar approximation for gluon loop contribution reads
\begin{align}\label{Pi_gluon_approx1}
&P_T^{m'm}(Q)\P^{0m<(1)}_{g,hard}\simeq-2\fctrg\int_0^\infty k^2dk\frac{1}{2\p}2kq i\e^{lm'a}\o_a^\pp\frac{k}{2q}\frac{q_0}{q}\frac{1}{2kq}\frac{1}{2k}f'(k),\nonumber\\
&P_T^{m'm}(Q)P_T^{n'n}\P^{mn<(1)}_{g,hard}\simeq-2\fctrg\int_0^\infty k^2dk\frac{1}{2\p}(-1)i\e^{n'm'a}\o_a^\pr k^2\(1-\frac{q_0^2}{q^2}\)\frac{1}{2kq}\frac{1}{2k}f'(k),
\end{align}
with the subscript ``g,hard'' denoting gluon loop contribution from hard loop momenta.
\eqref{Pi_gluon_approx1} also scales as $T^2\o/q$. However, it is incomplete. Because of Bose enhancement, there is an additional contribution from the phase space with $K\sim Q$. In this case, we may expand the distribution function $f(k_0+q_0)\simeq \frac{T}{k_0+q_0}-\frac{1}{2}$, $f'(k_0)\simeq -\frac{T}{k_0^2}$ to obtain
\begin{align}\label{ksimq}
&P_T^{m'm}(Q)\P^{0m<(1)}_{g,soft}\simeq-\fctrg\int_{\frac{q\mp q_0}{2}}k^2dk\frac{1}{2\p}(2k+q)_02q\big[\(2-\frac{(k_\pr+q)q}{(k_0+q_0)^2}\)\frac{k_\pp^2}{4k_0^2}\nonumber\\
&+\(2+\frac{k_\pr}{2q}-\frac{k_\pp^2+2(k_\pr+q)^2}{4(k_0+q_0)^2}\)\frac{1}{2}\big]\frac{1}{2kq}\frac{1}{2k}\(\frac{T}{k_0+q_0}-\frac{1}{2}\)\frac{T}{k_0^2}i\e^{lm'a}\o_a^\pp\hat{q}_l|_{k_0=\pm k},\nonumber\\
&P_T^{m'm}P_T^{n'n}\P^{mn<(1)}_{g,soft}\simeq-\fctrg\int_{\frac{q\mp q_0}{2}}k^2dk\frac{1}{2\p}\big[\(-\frac{k_\pp^2k_\pr q}{2k_0^2}+\frac{k_\pp^2}{4}\)\(-\frac{4(k_\pr+q)q}{(k_0+q_0)^2}-4\)+4q\frac{k_\pp^2 k_\pr}{(k_0+q_0)^2}\frac{k_\pp^2}{2k_0^2}\nonumber\\
&+4\frac{k_\pr q k_\pp^2}{(k_0+q_0)^2}\(\frac{k_\pr^2}{2k_0^2}+\frac{1}{2}\)\big]\(\frac{T}{k_0+q_0}-\frac{1}{2}\)\frac{T}{k_0^2}i\e^{n'm'a}\o_a^\pr|_{k_0=\pm k},
\end{align}
with the subscript ``g,soft'' denoting gluon loop contribution from soft loop momenta.
There are two contributions in \eqref{ksimq} coming from the poles at $k_0=\pm k$, with the corresponding integration lower bound set by $\frac{q\mp q_0}{2}$ respectively. We will extract the leading $T^2\o/q$ contribution, which obviously comes from the term $\frac{T}{k_0+q_0}$ in the expansion of $f(k_0+q_0)$. The corresponding integral can be calculated analytically. While the contributions from the poles at $k_0=\pm k$ are separately ultraviolet (UV) divergent, their sum is UV safe\footnote{The UV safe property is not true for non-leading terms like $T\o$, for which a matching of hard and soft parts is needed.}. Pushing the integration upper bound to $\infty$, we obtain
\begin{align}\label{soft}
&P_T^{m'm}(Q)\P^{0m<(1)}_{g,soft}=-\fctrg\frac{1}{2\p}\frac{T^2}{q}\frac{1}{48 q x^5 \left(x^4-10 x^2+9\right)}\big[-54 x + 123 x^3 + 725 x^5 - 447 x^7 + 177 x^9\nonumber\\
& - 12 x^{11} + 
12 x^5 (18 - 2 x^2 - 9 x^4 - 8 x^6 + x^8) \tanh^{-1}\frac{1-x}{2} + 
30 x^6 (9 - x^2 - 9 x^4 + x^6) \tanh^{-1}\frac{x}{3}\nonumber\\
& + 3 (-3 + x) (-1 + x) (1 + x) (3 + 
x) ((2 - 3 x^2 + 28 x^4 - 13 x^6 - 10 x^8) \tanh^{-1}x + 
4 x^5 (2 + 2 x^2 + x^4)\nonumber\\
& \times\tanh^{-1}\frac{1+x}{2})\big]i\e^{lm'a}\o_a^\pp\hat{q}_l,\nonumber\\
&P_T^{m'm}P_T^{n'n}\P^{mn<(1)}_{g,soft}=-\fctrg\frac{1}{2\p}\frac{T^2}{q}\frac{1}{24x^4}\big[-30 x^7+35 x^5-16 x^3+3 (x-1) (x+1) \left(10 x^6-5 x^4+6 x^2+17\right) \nonumber\\
&\times\tanh ^{-1}x+51 x\big]i\e^{n'm'a}\o_a^\pr.
\end{align}
The hard parts for quark \eqref{Pi_quark_approx} and gluon \eqref{Pi_gluon_approx1} can be calculated and combined as
\begin{align}\label{hard}
&P_T^{m'm}(Q)\(\P^{0m<(1)}_{g,hard}+\P^{0m<(1)}_{q}\)=-\frac{1}{2\p}\frac{\p^2g^2T^2x}{6q}\(N_f+\frac{C_A}{2}\)i\e^{m'al}\o_a^\pp\hat{q}_l,\nonumber\\
&P_T^{m'm}P_T^{n'n}\(\P^{mn<(1)}_{g,hard}+\P^{mn<(1)}_{q}\)=-\frac{1}{2\p}\frac{\p^2g^2T^2(1-x^2)}{6q}\(N_f+\frac{C_A}{2}\)i\e^{n'm'a}\o_a^\pr.
\end{align}
We have verified that \eqref{soft} and \eqref{hard} combined accurately reproduces direct numerical integration of the full expressions when $q\ll T$. Moreover, the IR limit of $D_{m'n'}^{<(1)}$ and $D_{0m'}^{<(1)}$ given by \eqref{soft} and \eqref{hard} scale the same as the counterpart $D_{\m\n}^{<(0)}$ in the absence of vorticity.

%\begin{footnotesize}
%\begin{thebibliography}{999}%\refmark

%\end{thebibliography} %\end{footnotesize}
%\end{multicols}

\end{CJK}

\end{document}